\documentclass[acmsmall,screen]{acmart} % ,,review,anonymous
\usepackage{xspace}
\usepackage{verbatimbox}
\usepackage{algorithm}
\usepackage[noend]{algpseudocode}
\usepackage{wrapfig}
\usepackage{caption} 
\usepackage{subcaption}
\usepackage{realboxes}
\usepackage{xcolor}
\usepackage{color}
\usepackage{listings}

\usepackage[normalem]{ulem} 

\newcommand{\tool}{{\sc DuoGlot}\xspace}
\newcommand{\toolns}{{\sc DuoGlot}}
\newcommand{\toolbf}{{\sc \textbf{DuoGlot} \xspace}}
\newcommand{\toolbfns}{{\sc \textbf{DuoGlot}}}

\definecolor{darkspringgreen}{rgb}{0.09, 0.45, 0.27}
\definecolor{deletecolor}{rgb}{0.66, 0, 0}
\definecolor{addcolor}{rgb}{0, 0.66, 0}

\newcommand\todoa[1]{\textcolor{black}{#1}}

\newenvironment{todoenv}
    {\color{black}} %{\color{addcolor}}  % ===== change this =====
    {\color{black}}

\newcommand{\pylang}{{Python}\xspace}
\newcommand{\jslang}{{JavaScript}\xspace}

\newcommand{\initialsetcount}{44\xspace}
\newcommand{\totalrulecount}{142\xspace}
\newcommand{\rulecountsimple}{82\xspace}
\newcommand{\rulecountmedium}{9\xspace}
\newcommand{\rulecountcomplex}{7\xspace}

\newcommand{\snippetcc}{5389\xspace}
\newcommand{\snippetlc}{360\xspace}
\newcommand{\benchlc}{5297\xspace}

\newcommand{\codexrawacc}{35\%\xspace} 
\newcommand{\codexfixacc}{74\%\xspace} 
\newcommand{\jthonacc}{43\%\xspace} 
\newcommand{\pyjsacc}{21\%\xspace} 
\newcommand{\tscryptacc}{76\%\xspace} 
\newcommand{\transcoderjavatime}{1.36}
\newcommand{\transcodercpptime}{2.85}

\newcommand{\boot}{\initialsetcount\xspace}
\newcommand{\tot}{\totalrulecount\xspace}

\newcommand{\gfgbenchmark}{GeeksforGeeks\xspace}

\newcommand{\syntaxchecker}{syntax checker\xspace}

\newcommand{\mycentered}[1]{\begin{tabular}{l} #1  \end{tabular}}
\newcommand{\mycenteredbuf}[1]{\mycentered{#1}}  

\renewcommand{\arraystretch}{1.2}
\usepackage[skip=5pt]{caption}

\definecolor{mygray}{HTML}{e3e6e8}
\newcommand{\code}[1]{{\setlength\fboxsep{1pt}\colorbox{mygray}{\lstinline|#1|}}}
\newcommand{\codesh}[1]{{\setlength\fboxsep{1pt}\colorbox{mygray}{\lstinline|#1|}}}

\AtBeginDocument{%
  \providecommand\BibTeX{{%
    \normalfont B\kern-0.5em{\scshape i\kern-0.25em b}\kern-0.8em\TeX}}}

\setcopyright{none}
\usepackage{pgfplotstable}

\pgfplotsset{
    /pgfplots/area cycle list/.style={
        /pgfplots/cycle list={%
            {purple,mark=o,mark size=2pt,fill=blue!30!white},%
            {blue,fill=red!30!white,mark=*},%
            {yellow!60!black,fill=yellow!30!white,mark=*},%
            {black,fill=gray,mark=*},
        }
    },
}

\pgfplotstableread[col sep=comma]{data/fig-rulecountperfcalc.csv}{\loadedtable}
\pgfplotstableread[col sep=comma]{data/fig-rulecountapp.csv}{\loadedtableapp}
\pgfplotstableread[col sep=comma]{data/fig-rulecountused.csv}{\loadedtableused}
\begin{document}

\title{User-Customizable Transpilation for Scripting Languages}

\author{Bo Wang}
\affiliation{%
  \institution{National University of Singapore}
  \country{Singapore}
}

\author{Aashish Kolluri}
\affiliation{%
  \institution{National University of Singapore}
  \country{Singapore}
}

\author{Ivica Nikoli\'{c}}
\affiliation{%
  \institution{National University of Singapore}
  \country{Singapore}
}

\author{Teodora Baluta}
\affiliation{%
  \institution{National University of Singapore}
  \country{Singapore}
}

\author{Prateek Saxena}
\affiliation{%
  \institution{National University of Singapore}
  \country{Singapore}
}
\begin{abstract}
  A transpiler converts code from one programming language to another. 
  Many practical uses of transpilers require the user to be able to guide or {\em customize} the program produced from a given input program. This customizability is important for satisfying many application-specific goals for the produced code such as ensuring performance, readability, ease of exposition or maintainability, compatibility with external environment or analysis tools, and so on. Conventional transpilers are deterministic rule-driven systems often written without offering  customizability per user and per program. Recent advances in transpilers based on neural networks offer some customizability to users, e.g. through interactive prompts, but they are still difficult to precisely control the production of a desired output. Both conventional and neural transpilation also suffer from the ``last mile'' problem: they produce correct code on {\em average}, i.e., on most parts of a given program, but not necessarily for all parts of it.    
  We propose a new transpilation approach that offers fine-grained customizability and reusability of transpilation rules created by others, without burdening the user to understand the global semantics of the given source program. Our approach is {\em mostly automatic} and {\em incremental}, i.e., constructs translation rules needed to transpile the given program as per the user's guidance piece-by-piece. Users can rely on existing transpilation rules to translate most of the program correctly while focusing their effort locally, only on parts that are incorrect or need customization. This improves the correctness of the end result.
\todoa{
We implement the transpiler as a tool called \tool, which translates Python to Javascript programs, and evaluate it on the popular \gfgbenchmark benchmarks. 
\tool achieves $90\%$ translation accuracy and so it outperforms all existing translators (both handcrafted and neural-based), 
while it produces readable code. 
We evaluate \tool on two additional benchmarks, containing more challenging and longer programs,  
and similarly observe improved accuracy compared to the other transpilers. 
}

\end{abstract}

\maketitle

\section{Introduction}
\label{sec:intro}

Transpilers are designed to automatically translate source code written in one programming language to its equivalent code in another. They have many applications including migrating codebases to newer languages and compiling code written in a newly-designed language to a well-supported language. To illustrate, the Common Bank of Australia is reported to have spent $750$ million dollars over $5$ years to replace its codebase written in COBOL~\cite{linkcobol}. This problem is exacerbated when languages are frequently updated to newer versions that selectively break backward compatibility, such as in the case of \pylang v2 to v3~\cite{py3incompatible} and Tensorflow v1 to v2~\cite{tf2migrate}. In all of these situations, taking assistance from transpilers could improve productivity and reduce costs. Transpilers have other applications, such as in education, where students can learn new programming languages by writing example code in languages they may be familiar with~\cite{pasternak2017tips}.

Several ways of designing transpilers have been proposed. The conventional way is for human experts to manually write transpilers as a collection of translation rules~\cite{cordy2006txl}. 
On the other hand, 
neural machine translation (NMT) is used to learn translation patterns or models from a large corpus of code from source and target languages---a promising approach to reduce human effort in writing transpilers~\cite{xinyun2018tree,roziere2020unsupervised,roziere2021leveraging,mariano2022automated}. All existing approaches suffer from at least one of two major drawbacks. The first is related to the {\em monolithic costs} required for development of transpilers.  It takes considerable time and expertise to write translation rules that are sufficiently complete to transpile a significant fraction of programs in the source language for conventional transpilers. The upfront costs of designing and integrating such monolithic transpilers into the development pipelines can outweigh the benefits. Often users are  interested in translating only certain programs and are not willing to invest the effort needed to write general transpilers.
NMT transpilers also incur large monolithic costs. They require hundreds of thousands lines of code in the source and target languages as examples,  access to large general language models such as GPT-3/Codex~\cite{brown2020language,chen2021evaluating}, and are computationally expensive to train.
Without sufficient training data and computational costs incurred, neural machine translators can have low transpilation accuracy.
The second problem is that of {\em inflexibility}. Existing transpilers assume that all users prefer the same translations of the same source program. There are multiple ways to translate the same code and developers often have many soft constraints to satisfy (library dependencies, preferred paradigms, readability, performance, etc) when choosing a specific way of translation. Often, there is a gap between the user's intent and the translation produced by existing transpilers because they have been designed to work in generic contexts rather than the one the user wants. 

\noindent \textbf{Our approach.} In this paper, we propose a new approach to designing transpilers that tackle the above drawbacks. We advocate that transpilers should treat {\em user-customizability} as a first-class design objective. In our proposed approach, users can selectively write or reuse translation rules that are sufficient to transpile their specific program of interest correctly. The power of the approach comes from two sub-features. The rules are created {\em incrementally} and {\em mostly automatically}. By incremental, we mean that the user can select the starting translation rule set, potentially written by others, and incrementally add to it until the given program is correctly translated. By mostly automatic, we mean that the transpiler tries to minimize the user effort by learning the transpilation rules from code snippets given as ``hints'' by the user. The process should be data efficient, i.e., learning a new rule should require very few pairs of source-target code snippets as examples.

We build a prototype system called \tool to transpile programs between two popular scripting languages: \pylang to \jslang. We choose \pylang and \jslang for illustration of the broader concept and its practical feasibility, though we expect supporting other untyped scripting languages to be similarly feasible.
\tool takes as input a \pylang program and produces as output a \jslang program. The goal is to produce a correctly translated program, i.e.,  to generate target code with the same input-output behavior as the source program under given unit tests. Several design decisions have made this approach practical. First, \tool translates the abstract-syntax tree (AST) representation of one program to its AST representation in another language, similar to some of the existing transpilers. Converting code to AST using standard compilers and vice versa using pretty-printers is practical as tools for this are commonly available for many popular languages. Second, our transpiler is a ``stateless'' transducer. 
\todoa{By avoiding explicit state transition rules, \tool reduces overfitting and makes rules learned from code examples more usable,}
i.e., the rules created with \tool can be applied to a similar statement elsewhere in the same program, or imported when translating a similar benchmark or a library. % 
Lastly, the transducer is designed to permit non-determinism, i.e., it offers several choices at a given point during translation. This is key to making the transpilation adaptive to the program being translated, as it is forced to pick between translation rules dynamically.
Lastly, all rules created are open to inspection by the user, making the approach more interpretable compared to NMT.

We demonstrate \tool's performance by 
translating \pylang code to \jslang \todoa{on one 
existing and two new benchmarks.
}
\todoa{
First, we evaluate on
the popular benchmarks maintained by the} TransCoder project~\cite{roziere2020unsupervised,roziere2021leveraging,linkgfg}. 
The benchmarks have $699$ \pylang programs (1-40 lines) with unit tests.
We compare \tool to $3$ manually written rule-based transpilers available publicly and to Codex, a modern neural-based transpiler that supports \pylang to \jslang. \tool translates $90\%$ of the benchmarks with just $\totalrulecount$ rules learned from pairs of code snippets (one or two for each rule) provided by the user. All other evaluated tools correctly translate at least $14\%$ fewer programs. 
Furthermore, as an illustration of how the rules can be customized for soft constraints, we measure the readability of the transpiled code using the code bloat metric \todoa{(an increase in the number of characters)}. Less code bloat is better~\cite{alkhatib1992maintenance}. 
\tool produces code that has good readability, comparable to that of Codex, and $1.6\times$ to $8\times$ better than the hand-crafted transpilers we study. 
Qualitatively, we analyze the manual effort involved in working with \tool, providing examples and a rough categorization. We argue that most of the rules are rather simple and so are the required code snippets. 

\todoa{We further assess \tool's strengths and limitations with two cases studies, each focused on running the tool on new benchmarks.  
The first case study is about translating more authentic and complex programs, with potentially insufficient unit tests. We collect solutions to LeetCode programming challenges\footnote{\url{https://leetcode.com/problemset/all/}} written in Python (each has a few unit tests in their problem description) and translate them to JavaScript.
We end up with 1067 such Python programs, with an average length of 15 lines. 
Due to increased program complexity, no other transpiler is able to achieve accuracy higher than 50\%.  
\tool achieves 75.4\% accuracy with 106 additional on top of the previous 142 rules, and requires 1.75 seconds per program. 
In the second case study we test the performance on longer code. We collect benchmarks from the book ``Cracking the coding interview''\cite{mcdowell2015cracking} -- in total 25 Python programs, each with  
54 to 160 lines of code. 
Once again, \tool has the best accuracy among the transpilers, while increasing its average translation time to 6.8 seconds per program. It achieves 56\% accuracy with 197 rules (142 previous and 55 additional), which is higher than the accuracy of the other transpilers (12\% by Codex and 44\% by Transcrypt). 
The two case studies reveal some of the limitations of \tool as well. First, despite being more correct than other transpilers, \tool falls short of the ideal 100\% accuracy. Second, when the unit tests do not provide full coverage, \tool may have unpredictable accuracy on the uncovered code. Third, \tool is scalable, but with some degradation of accuracy and efficiency. 
Despite being mostly generic, the above limitations indicate that \tool may be improved, and we leave this as an open problem. 
}

\section{Problem Setup and \toolbfns's Overview}
\label{sec:overview}
\begin{figure}[!htbp]
    \centering
    \begin{tabular}{@{}c@{}}
            \includegraphics[width=0.8\textwidth]{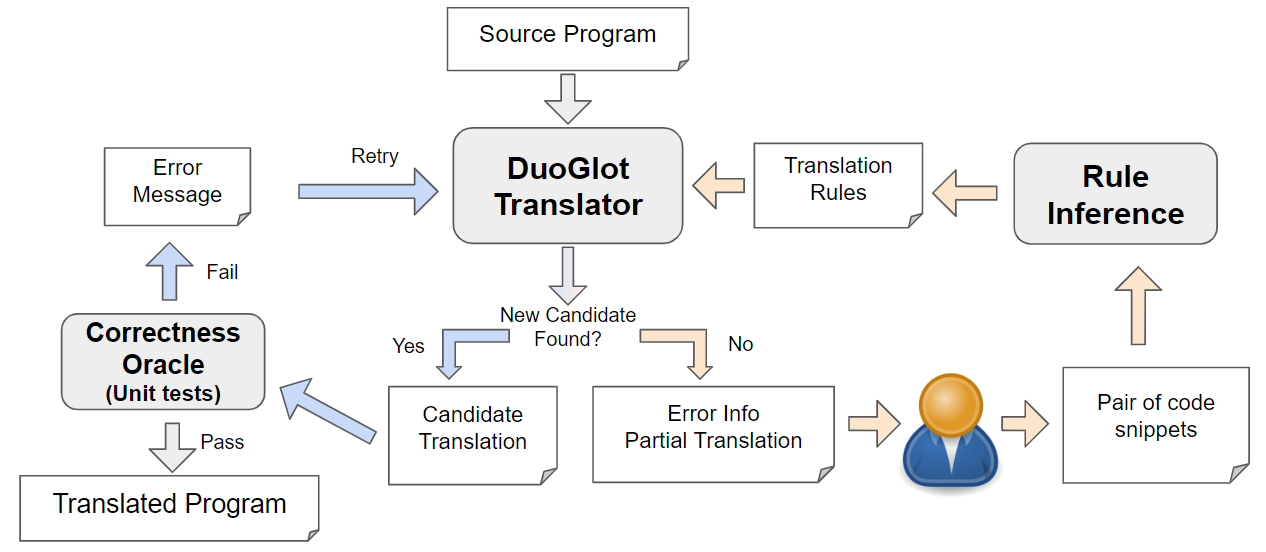}\\ 
        \small  \\
    \end{tabular}
    \caption{Workflow of Incremental Translation. The loop in blue arrows on the left is the machine loop (frequent case) and the loop in yellow arrows on the right is the user loop (less frequent case, incremental).}
    \label{fig:motivateflow}
\end{figure}

A user utilizes the transpiler to translate a program written in  source language to a program written in  target language. We design a new transpiler called \tool that translates programs from \pylang (source) to \jslang (target).   We assume that unit tests for both the source and the target programs are available to validate the correctness of the translation. 
\todoa{
Test-based validation is used in practice, for instance in programming contests like LeetCode.
If there are unit tests only for the source code, it is relatively easy to translate them  to the target language as demonstrated previously~\cite{roziere2020unsupervised}.}
When transpilers cannot translate a particular source program they usually abort or output a wrong target program. 
An interesting alternative, however, would be to infer a new rule that handles the problematic part of the program by prompting users for a fix. This is the approach we take in \tool by asking users for code snippets in both languages as hints for fixing the translation. 
The paradigm of interacting with users to achieve the required functionality is used in program synthesis~\cite{gulwani2017program} and verification~\cite{bader2018gradual}. 
For instance, popular programming-by-example synthesis tools, such as Flashfill~\cite{polozov2015flashmeta}, 
receive input-output examples from the user to make the search for a correct program tractable. 

\begin{wrapfigure}{r}{0.48\textwidth}
    \centering
    \includegraphics[width=0.47\textwidth]{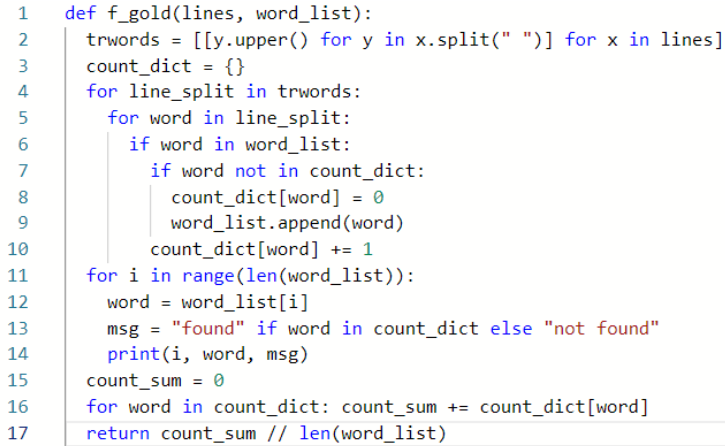}
    \caption{Python program to translate.}
    \label{fig:pythoncode}
    \vspace{5pt}
    \includegraphics[width=0.47\textwidth]{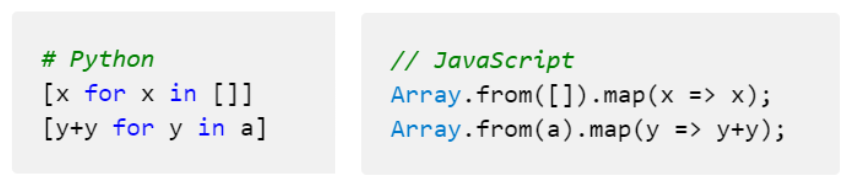}
    \caption{\todoa{Pairs of source-target code snippets} for translation of list comprehension.}
    \label{fig:codesnippets}
    \vspace{5pt}
    \includegraphics[width=0.47\textwidth]{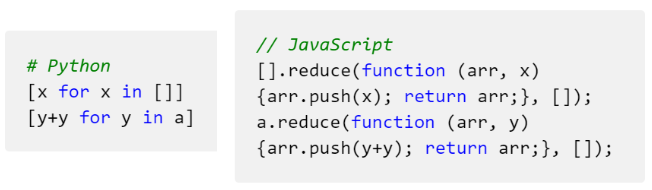}
    \caption{\todoa{Pairs of source-target code snippets} for alternative translation of list comprehension.}
    \label{fig:codes-alt-nippets}    
\end{wrapfigure}

The workflow for our transpiler \tool is shown in Figure~\ref{fig:motivateflow}. Our transpiler puts the user in control of the transpilation process. \tool is an incremental transpiler. It tries to learn a new rule each time it cannot translate some part of the code and thus slowly grows its translation \emph{rule set}. To learn each such new rule, it interacts with the user and prompts for pairs of code snippets in source and target languages that help to infer the rule. The rule focuses the user's attention locally on one part of the code, so the user does not need to understand the global semantics of the source program.
These user-provided snippets aid rule inference, while letting the user customize the translation directly with the provided snippets. Note the user is not expected to be an expert---they do not need to learn a third language or know the internals of \tool.

\todoa{At the beginning, \tool starts the translation of the source program} with a set of general translation rules to bootstrap the translation procedure. These rules capture the most prominent constructs of the languages such as identifiers, literals, basic assignments, \texttt{if} statements, and so on. 
All rules learned during previous translations are also added to the rule set and the user has complete control over which rules it selects. 
\tool then searches for a candidate source program translation using the provided rules. 
If the search concludes without finding a correct translation, 
\tool locates the place of code that failed to translate and prompts the user for help to handle the failed statement or expression. 
The user then provides short code snippets that demonstrate the correct translation and the preferred translation style for such a statement or expression. At this point, \tool executes its rule inference procedure to learn a rule, by looking at syntactic patterns of the provided snippets. It adds this newly learned rule to its rule set so that it can directly apply the learned rule when the same syntactic pattern appears elsewhere in the future. Using the updated rule set, \tool  retries the search for more candidate translations, until the translation task is finished.

\tool is {\em mostly} automatic. When the user provides code snippets as hints for a fix, \tool does much of the heavy lifting in learning the translation rule internally from the given hints. Further, during the transpilation process as new candidate translations get created, \tool incrementally checks their correctness by running unit tests. If the translation turns out to be incorrect then an error message with the error location (when available) is automatically fed back. Hence the correctness oracle  not only validates but also serves as a test-repair loop. This helps to generate correct translations efficiently by pruning away bad translations eagerly. To our knowledge,  existing transpiler designs lack such loops that enforce correct translations. 

\begin{wrapfigure}{r}{0.5\textwidth}
    \includegraphics[width=0.46\textwidth]{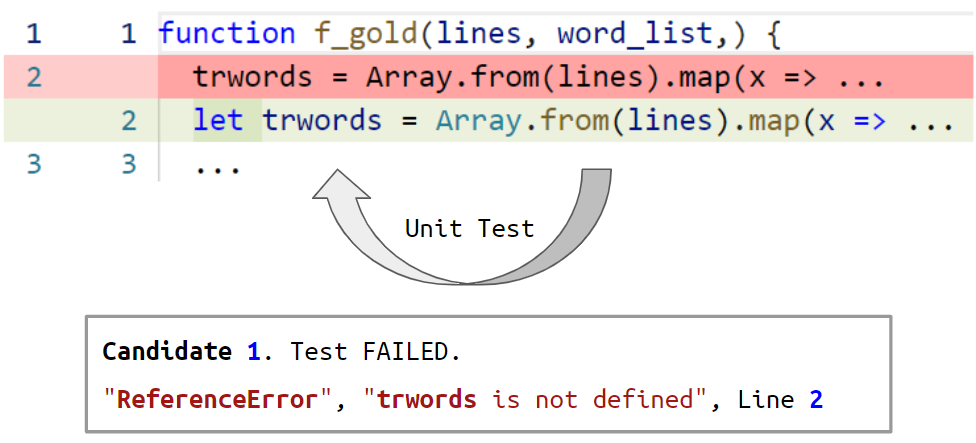}
    \caption{Candidate translation 1 has an error on line 2. \tool will fix such error automatically.}
    \label{fig:jsfix}
    \vspace{0pt}
    \includegraphics[width=0.46\textwidth]{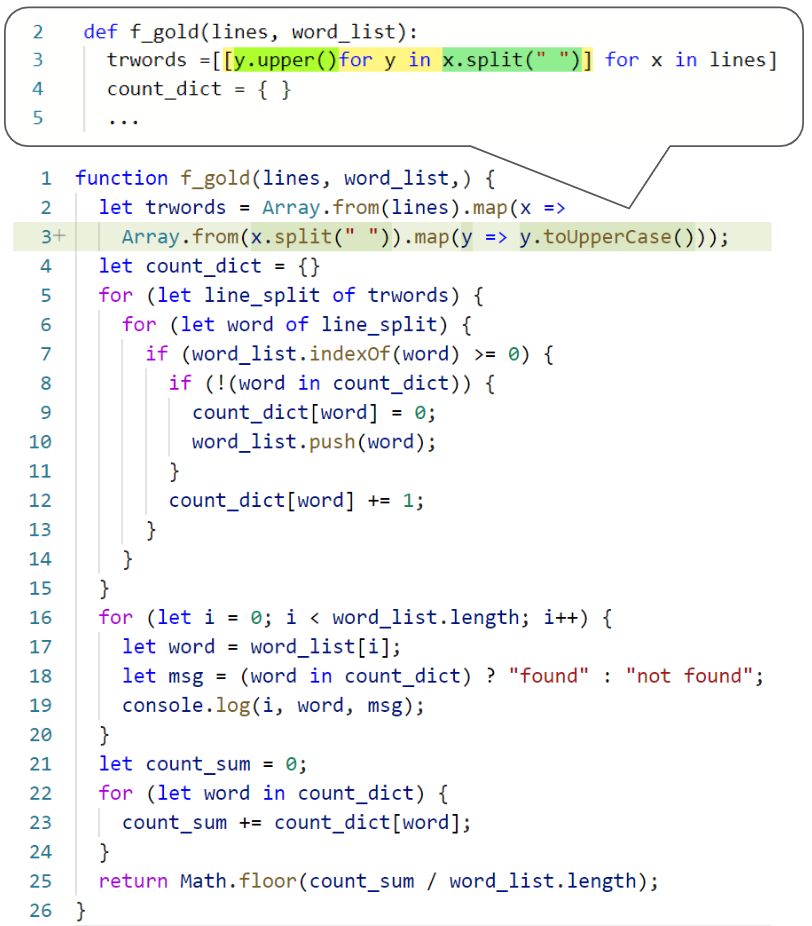}
    \caption{Translated program (Candidate 13). The mapping between pieces of code is shown in the UI.
    }
    \label{fig:jscode}
\end{wrapfigure}

\paragraph{Motivating Example}
\label{sec:motivation}
To illustrate the workflow of our transpiler, consider the Python program given in Figure~\ref{fig:pythoncode}. The program outputs the average frequency of a set of words as they appear in a set of lines. Assume we start \tool with a small rule set insufficient to translate the whole program. For instance, there is no rule to translate  \pylang's list comprehension used in line 2. 
Our tool detects this, notifies the user there is a problem at line $2$, and asks for code snippets. 
The user can reply with simple snippets like the ones in Figure~\ref{fig:codesnippets}. Note that list comprehension can be translated in different ways in \jslang, and the user may instead prefer and  provide the alternative style from Figure~\ref{fig:codes-alt-nippets}. (This is the   customization feature of the transpiler---the user determines the preferred style through the provided code.) 
Subsequently, \toolns's rule learning procedure automatically infers a translation rule from these code snippets and adds it to the rule set. The user, if desired, may directly inspect the rule.

Then \tool retries the translation and finds the candidate translation for the list comprehension. After applying this rule, it generates a candidate translation $1$  (the first two lines are shown in Figure~\ref{fig:jsfix}) where it assigns the list comprehension to the \code{trwords} variable, but without declaring the variable. Note, there are two competing rules in the bootstrapped rule set for a \pylang expression with \code{=} operator such as \code{a=b}. The first rule would keep the expression as is in the \jslang translation. The second rule would translate it to a declaration in \jslang, \code{let a=b}. The two rules arise since \pylang does not have a separate representation for assignments and declarations whereas \jslang does. However, the application of the first rule leads to the candidate translation $1$ which fails the tests and the correctness oracle reports that \code{trwords} at line $2$ is not defined. \tool will then look up its rule set and will figure out that it has another rule to apply at line 2 (which translates \pylang assignment to \jslang declaration). Thus, it will produce candidate translation $2$.

This iterative test-driven procedure will continue for $13$ rounds\footnote{It happens to be 13 rounds in this particular example.} in our example until a candidate translation that passes all the unit tests is found. Unlike in candidate $1$, in candidate $13$ all the syntax patterns are translated correctly including \codesh{for x in y} to \codesh{for (let x of y)}, \codesh{for (let x in y)} or \codesh{for(let i = 0; i < ..., i++)}. The final \jslang program is given in Figure~\ref{fig:jscode}. The translation is provided in the user interface (UI) where the user can hover on a piece of translated code to find out the part of source it comes from and the used translation rule.

\begin{figure}[!htbp]
    \centering
    \begin{tabular}{@{}c@{}}
            \includegraphics[width=\textwidth]{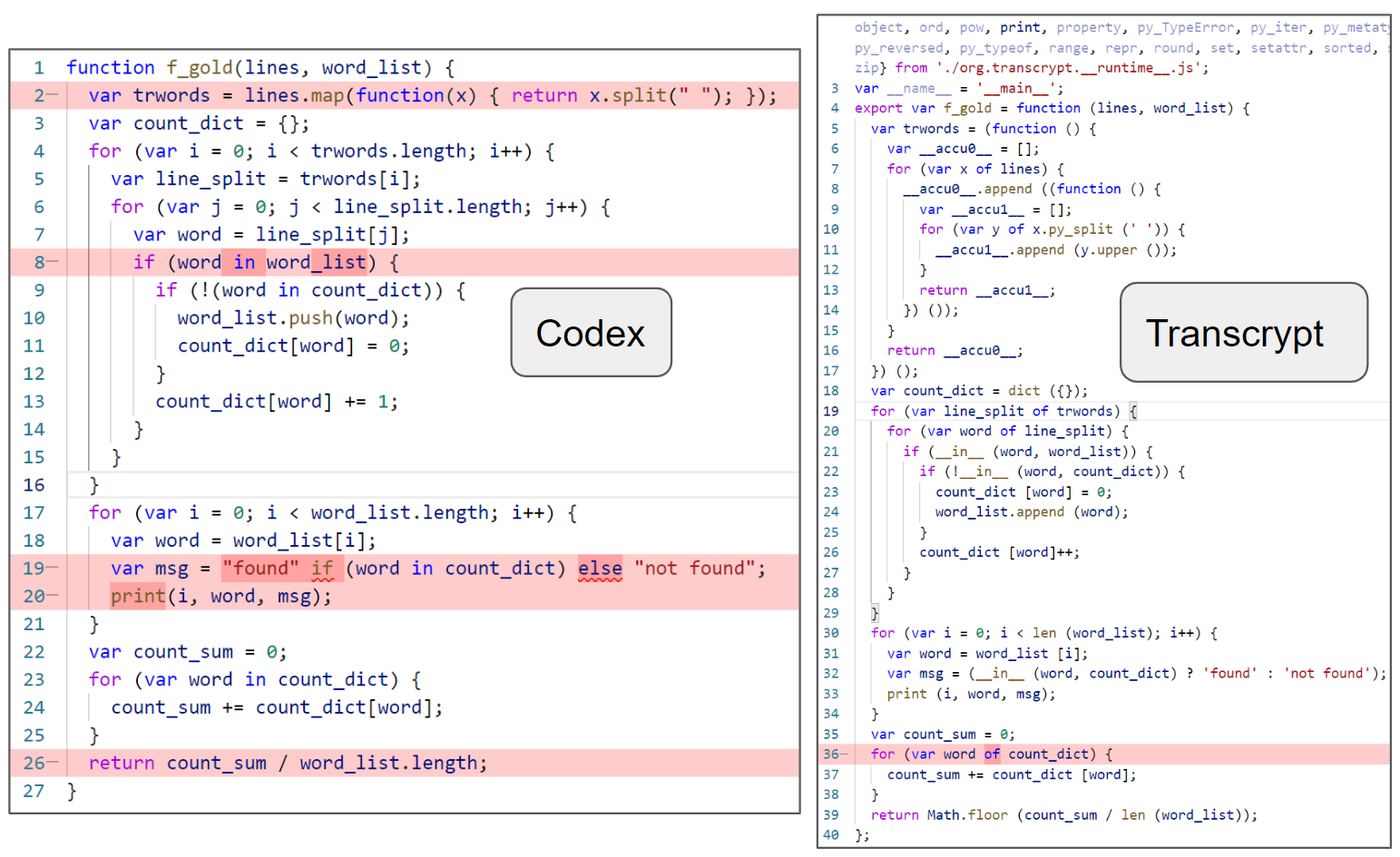}\\ 
        \small  \\
    \end{tabular}
    \caption{
    Codex's translation of the code in Figure~\ref{fig:pythoncode} has multiple errors, mostly due to confusing semantically similar operators and expressions. Transcrypt's translation emulates many \pylang's built-in functions, but still has one error due to the type difference between \texttt{Object} and \texttt{Array} and it is much more verbose.}
    \label{fig:motivatecompare}
\end{figure}

\paragraph{Comparison to Codex and Transcrypt}
To highlight the benefits of our approach, we also compare the translation of  Figure~\ref{fig:pythoncode} program to two state-of-the-art transpilers, Codex~\cite{chen2021evaluating} and Transcrypt \footnote{Python 3.7 to JavaScript compiler. \url{https://github.com/qquick/Transcrypt}} (see Figure~\ref{fig:motivatecompare} for their translated \jslang program). Codex is a machine-learning-based language model for code-related tasks used by Github Copilot, while Transcrypt is a popular hand-crafted rule-based transpiler reportedly used in production~\cite{linktranscryptinprod}. We observe that Codex frequently confuses semantically similar expressions, operators and variables, or misses out part of the source code. 
Let us consider the 5 errors in Codex translation highlighted in red. At line 2, Codex ignores the \code{toupper()} function call inside the nested list comprehension. 
At line 8, it translates the expression \code{word in word_list} as it is but \code{word_list} is not a dictionary and the correct translation is \code{word_list.indexOf(word) >= 0}. At line 19, the translation is not syntactically valid because it should be a ternary expression rather than an expression followed by a statement. At line 20, it translates \code{print} as it is, while 
at line 26, it confuses \code{//} with \code{/}. 
There is no easy way to predict whether and why Codex will make a mistake because there are {\em no} interpretable rules or explanations for its translation.
Overall Codex's output is readable, however, it makes frequent mistakes with semantically similar operators. A human working with Codex may be able to fix these issues, nonetheless, the Codex approach requires a huge training dataset and computational resources to arrive at this result.

Unlike Codex or \tool, Transcrypt imports a custom runtime library for emulating many common \pylang functions and operators, as shown at the top-right of Figure~\ref{fig:motivatecompare}. However, it still has an error at line $36$ where it fails to infer that \code{count_dict} is an object rather than an array, so \code{for (let x in y)} should be used instead of \code{for (let x of y)}. This highlights a key issue with designing a transpiler that aims to solve the general transpilation problem all at once: it is hard to accurately know the variable's type (or possible values), which is critical when translating to a different language in a different runtime. On the contrary, explicit type inference \todoa{is not even required} in the increment transpilation paradigm we present. The test-driven feedback loop of \tool  guides it towards the correct translation.

\vspace{10pt}
\noindent
In summary, our transpilation paradigm provides the following benefits: 
\begin{enumerate}
\item \textbf{Customizability}. 
\tool adopts user-provided translation rules.  
It can be customized for different coding styles, APIs, language versions, etc. Moreover, \tool  grants fine-grained customization while learning rules from code snippets, allowing users to specify the translation rules for all  language constructs.  It allows them to modify the translation rules and tinker with different rule sets by splitting or merging them. Moreover, the customization is optional, and users may choose to reuse the rules supplied by others, partially or in full. None of the existing transpilers provide such customizability, in part due to their non-incremental nature. The manual monolithic ones demand expert knowledge while the ML-based ones might need re-training.
\item \textbf{Low manual effort.} \tool \emph{automatically} learns the rules from small code snippets provided by the user. This alleviates the effort of manually crafting them. Further, \tool engages the user to provide code snippets only when learning a new rule, not already present in the knowledge base. Its incremental approach requires learning only rules needed to perform the translation task at hand. Therefore, by design, it requires a much smaller number of code snippets than, for instance, training a general ML model. The manual effort required post-translation is low as well since \tool is optimized to generate correct and readable code along with an explanation. In our experiments,  \tool obtains correct translations for $90\%$ of the main benchmarks, in total $\benchlc$ lines of code, 
with just $360$ lines of user-provided code snippets in total.
\end{enumerate}

\section{\toolbfns: Challenges and Approach}
\label{sec:approach}
The core component of \tool is a tree transducer (\cite{comon2008tree}, Chapter 6) which reads an input abstract syntax tree (AST) of the source program and translates it to the target AST. 
The target AST has enough information to be converted back to a syntactically-valid target program using a pretty printer.
This transducer can be seen as matching each sub-tree in the input AST and mapping it to a sub-tree in the target AST. The translation, therefore, proceeds as a sequence of steps where in each step the transducer uses a {\em translation rule} to output the translated AST. The transduction proceeds recursively top-down, i.e., it starts from the root of the input AST, applies the translation rule to the tree, and then moves down to translate some of its subtrees. Such transducers are called extended left-hand side (extended LHS) tree transducers~\cite{graehl2008training}. %,cite1,cite2,cite3
Translation rules are written in a domain-specific language (DSL), which we will detail later in Section~\ref{sec:transpacerule}. 
\tool addresses three key challenges to make the transpilation practical.

First, \tool needs to learn rules that are general and reusable, therefore avoiding overfitting to the code samples provided by the user. Suppose the code samples 
\code{a = 1; a = 2}
(\pylang) and 
\code{let a = 1; a = 2} 
(\jslang) are provided to learn a rule that handles assignment. \tool should not learn a rule that says ``assign value to a variable only after it is declared {\em and assigned the value 1}''. This is one form of overfitting and many such examples exist. To address this challenge, \tool's extended LHS tree transducer uses only a {\em single state}\footnote{For brevity, from now on we omit ``extended LHS'' when describing our tree transducer.}. 
Such a transducer
does not preserve any context or history across the sequence of translation steps, i.e., it applies the same translation to a sub-tree of the input AST regardless of how other sub-trees in the input AST have been translated.
So, \tool is forced to use the same translation for, say \code{a=2}, irrespective of how \code{a=1} was translated earlier. Therefore, it cannot have a rule that says ``translate \pylang assignment sub-expressions to \jslang assignment {\em only if the variable has been declared with value 1} or something else otherwise''. This fundamentally addresses part of the overfitting problem that may occur due to learning a rule that overfits to the given code snippets.

The second challenge is that of handling ambiguity during transduction. There are two sources of ambiguity: 1) the transducer does not have enough history or context to pick between rules;
and 2) there are inherently more than one functionally correct mappings to translate a given code snippet from \pylang to \jslang. To address the first source of ambiguity, the transducer needs to be {\em non-deterministic}, i.e., for the same input AST sub-tree there can be more than one rule option for the transducer to eventually apply. The decision of 
which rule to use, however, is not going to be manually specified in the state machine in the transducer since the single-state property of the transducer prevents that. Instead, the transducer will be forced to dynamically try out different rules at each step, and repeatedly will do so until it finds the choice that works.

\begin{figure}[!t]
    \centering
    \includegraphics[width=\linewidth]{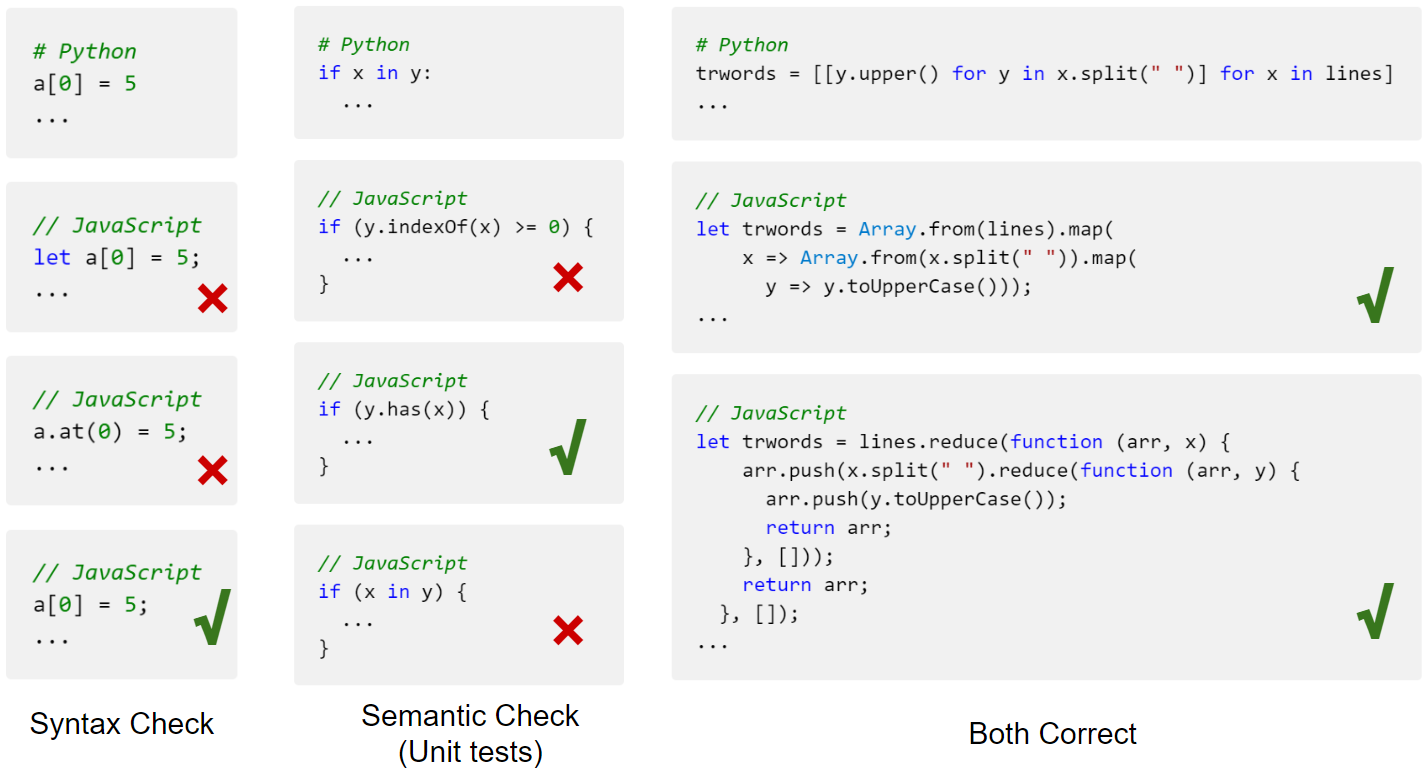}
    \caption{(Left) Syntactically valid and invalid translations. \tool checks the validity of every transduction step and rejects the invalid translations on the fly. (Middle) Syntactically-valid, but not necessarily semantically valid translations. The second translation is deemed correct when the dynamic type of \code{y} is a set. \tool checks for semantic validity by running the unit tests every time a syntactically-valid candidate translation is generated. (Right) Syntactically and semantically valid translations. In such a case, \tool just selects one based on pre-determined order of rules that is chosen arbitrarily to minimize the user effort.}
    \label{fig:examples}
\end{figure}

Resorting to non-determinism by itself may lead to combinatorial explosion as there are exponentially many rule combinations to be tried---several options exist at each step of the top-down traversal of the transducer on the input AST. The key observation is that among the many syntactic candidates at each step of the traversal, there are only a few syntactically correct choices and even fewer that eventually pass the given tests. Therefore, \tool eagerly validates the candidates using a syntax checker at each step,  and a semantic correctness checker that uses unit tests after a sequence of steps. This {\em incremental} validation significantly reduces the number of valid translation rule candidates that remain after a few steps of transduction.
For example, a transduction step translating \code{a[0] = 5} to \code{let . = .} followed by a transduction step translating \code{a[0]} to \code{.[.]} will be rejected eagerly by the syntax checker because \code{let a[0] = 5} is not syntactically valid. Similarly, the semantic correctness checker helps. The example in Figure~\ref{fig:examples} (middle) shows translation of the \pylang code \code{if x in y} to \jslang. There are three possible syntactically valid translations in \jslang: \code{y.indexOf(x) >= 0}, \code{y.has(x)}, and \code{x in y}. However, only one of them may be semantically correct, depending on  dynamic type of \code{y} (which needs to be a set). \tool's semantic checker implements a test-retry procedure to filter out such invalid translation choices. The procedure is invoked every time a syntactically valid candidate translation is generated so that it can run the unit tests to verify the correctness. Therefore, the transducer will produce a full translation by choosing the first option and by running the code with unit tests. If the runtime type of \code{y} is not an array, then an error from that line will be fed back to \tool so that it can immediately move to  the next option. 
\todoa{We emphasize that this is done automatically, without any user interaction, and on average requires from one to a few seconds to repair a dozen of errors.}

The second type of ambiguity (multiple functionally correct translations) is illustrated in Figure~\ref{fig:examples} (right). Such ambiguity can be resolved in two ways: ask the user \todoa{which rule they prefer to apply}, or choose the rules according to an arbitrarily pre-defined order. \tool allows both, and to minimize user interaction, by default it uses the latter. Specifically, it picks translation rules at each transduction step in a pre-determined order and moves forward until it either translates the whole program to a semantically correct translation or it gets stuck and backtracks. 

\begin{figure}[!ht]
    \centering
    \begin{tabular}{@{}c@{}}
            \includegraphics[width=\textwidth]{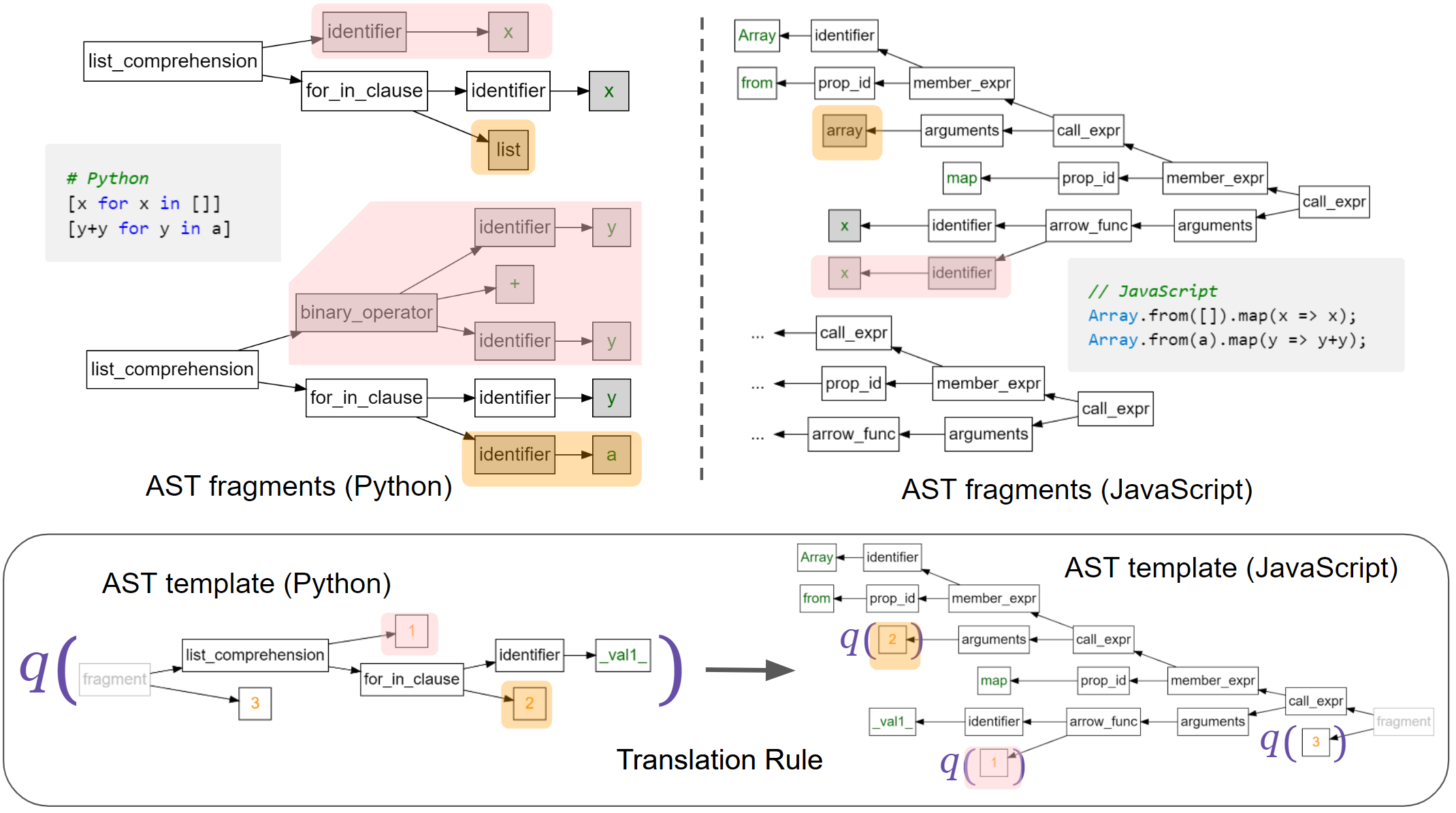}\\ 
            \includegraphics[width=\textwidth]{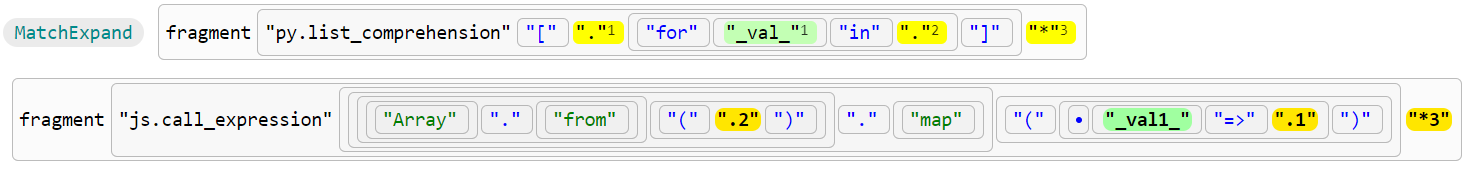}\\ 
    \end{tabular}
    \caption{The translation rule for translating list comprehension (bottom) is inferred from the AST fragments obtained from user-provided Python (left) and JavaScript (right) code snippets. The grey-shaded nodes are the non-common subtrees in the AST fragments, while the unshaded ones are the common ones. The pink-shaded subtrees are matched to the same hole (\texttt{1}) in the resulting AST template for each respective language. Similarly, the yellow-shaded subtrees are matched to same hole in their respective ASTs (\texttt{2}). \todoa{q(...) refers to subtrees of the source AST to be transduced.}}
    \label{fig:ruleinfer}
\end{figure}

When \tool has no valid rule left to try to translate the input AST, it interacts with the user to help derive a rule. It asks the user to provide code snippets in source and target languages. The third challenge \tool addresses is how to infer new rules fully automatically or with a minimal manual effort. 
For instance, in 
Figures \ref{fig:codesnippets} and \ref{fig:codes-alt-nippets} we show 
user-provided replies to a rule inference query for line $2$ of Figure~\ref{fig:pythoncode}. 
It is critical to understand that the query is localized at a single context in the source program, i.e., \tool prompts the user to specify how to translate one construct found in the code. \todoa{The user can provide snippets easily} because it is not necessary to understand the global context of the construct's use in the program. 

\tool automatically infers the translation rules by matching patterns observed in the ASTs of code snippets. The procedure is deterministic and works well in practice. Figure \ref{fig:ruleinfer} shows an example of how 
\tool's rule inference engine learns a translation rule from the snippets illustrated in Figure~\ref{fig:codesnippets}. 
Recall that in our running example we have two list comprehensions in Python for which we want to obtain the corresponding JavaScript translations. 
Briefly, \tool identifies an AST template that summarizes the two AST fragments corresponding to the list comprehensions in the \pylang snippets. The learned AST template consists of nodes that are common to both the AST fragments (not shaded boxes in Figure~\ref{fig:ruleinfer}) and holes that abstract away the uncommon parts between the AST fragments (grey-shaded boxes in Figure~\ref{fig:ruleinfer}).

The same procedure creates the AST template for the \jslang snippets provided, as shown in the right part of Figure~\ref{fig:ruleinfer}. 
The AST templates' holes are represented by numbers in the bottom part of the figure.
The translation rule is then simply the transform of the \pylang AST template to \jslang AST template, i.e., the fixed AST nodes of the source AST will be translated to the fixed  nodes of the target AST. The translation will leave the \pylang fragments matching the holes to be translated in the subsequent steps of the transducer.
We leave the details of how the AST templates are created and the handling of holes to Section~\ref{sec:components}.
Figure~\ref{fig:ruleinfer} also shows the learned rule in our DSL which can be inspected and optionally edited by the user to directly customize the inferred rule.

\section{\toolbfns: Key Components}
\label{sec:components}

\tool's four key components, the transducer (Section~\ref{sec:transpacerule}), the syntax and semantic checkers (Section~\ref{sec:validatept}), and the rule inference procedure (Section~\ref{sec:inferrule}) allow efficient translation of a program by incrementally learning the required rules from small code snippets. All of the key components are designed to work with any pair of dynamically typed scripting languages and \tool demonstrates this for \pylang and \jslang. 

\subsection{Non-Deterministic Single-State Tree Transducer}

\label{sec:transpacerule}

The transducer translates different parts of the source AST recursively by applying the translation rules. We design a domain-specific language (DSL) for writing the code translation rules, as shown in Figure~\ref{fig:transruledsl}. 
\tool requires a set of translation rules, which corresponds to the start symbol \textbf{Ruleset}. The \textbf{Ruleset} is a sequence of rules where each rule has a source and target pattern. The source pattern (\textbf{SrcPattern}) matches a source AST fragment which could be the whole tree or a subtree of the source AST. Similarly, the target pattern (\textbf{TrgPattern}) 
matches a target AST fragment. Therefore, when a source AST fragment matches with the source pattern in the current application of \textbf{Rule}, it can be translated into the corresponding target AST using the target pattern. Each pattern is a sequence of one or more \textbf{NodeMatcher}s which describes the ``nodes'' in the corresponding AST fragment. A node can be a non-terminal symbol (such as \texttt{forStatement} and \texttt{ifStatement}), a terminal symbol (any string), or another fragment altogether. Therefore, the fragment present in the current rule can be translated recursively using the rule that applies for that fragment. The recursive fragments are denoted by one of the placeholders \texttt{``.''} and \texttt{``*''}. The former matches a single arbitrary AST node representing a non-terminal symbol whereas the latter represents a sequence of AST nodes including non-terminal and terminal symbols. In the source pattern, these placeholders are represented by the \textbf{Capture} rule, whereas in the target pattern by the \textbf{Reference} rule. Each \textbf{Reference} in the target pattern corresponds to one of the \textbf{Capture}s in the source pattern. Therefore, the \textbf{Reference}s are just \textbf{Capture}s that are left as-is by the current rule until the next rule can proceed to match and expand (translate) them.

\begin{wrapfigure}{r}{0.53\textwidth}
    \renewcommand{\arraystretch}{0.55}
    \centering
    \footnotesize
    \begin{tabular}{@{}r@{\hskip3pt}c@{\hskip3pt}p{6cm}@{}}
        \textbf{Ruleset} & ::= & \textbf{Rule}+ \\
        \textbf{Rule} & ::= & (MatchExpand \textbf{SrcPattern} \textbf{TrgPattern})\\
        \textbf{SrcPattern} & ::= & (fragment \textbf{NodeMatcher}+)\\
        \textbf{TrgPattern} & ::= & (fragment \textbf{NodeExpander}+)\\
        \textbf{NodeMatcher} & ::= & \textbf{NTMatcher} | \textbf{TMatcher} | \\                         &    & \textbf{NTMatcherTpl} | \textbf{TMatcherTpl} |\\
                   &    & \textbf{Capture}\\
        \textbf{NodeExpander} & ::= & \textbf{NTExpander} | \textbf{TExpander} |\\
                            &     & \textbf{NTExpanderTpl} | \textbf{TExpanderTpl} |\\
                            &     & \textbf{Reference}\\

        \textbf{NTMatcher} & ::= & (\textbf{NTSymbol} \textbf{NodeMatcher}+)\\
          \textbf{NTExpander} & ::= & (\textbf{NTSymbol} \textbf{NodeExpander}+)\\
        
        \textbf{TMatcher} & ::= & (str \textbf{String}) | (nostr) | (val \textbf{String})\\
          \textbf{TExpander} & ::= & (str \textbf{String}) | (nostr) | (val \textbf{String})\\
        
        \textbf{NTMatcherTpl} & ::= & (``\_nt\_'' \textbf{NodeMatcher}+)\\
          \textbf{NTExpanderTpl} & ::= & (``\_nt\textbf{Idx}\_'' \textbf{NodeExpander}+)\\
          
        \textbf{TMatcherTpl} & ::= & ``\_str\_'' | ``\_val\_''\\
          \textbf{TExpanderTpl} & ::= & ``\_str\textbf{Idx}\_'' | ``\_val\textbf{Idx}\_''\\
          
        \textbf{Capture} & ::= & ``.'' | ``*''\\
          \textbf{Reference} & ::= & ``.\textbf{Idx}'' | ``*\textbf{Idx}''\\
        \textbf{NTSymbol} & ::= & ``\textbf{Lang}.\textbf{NTName}''
        
    \end{tabular}
    \footnotesize
    \begin{tabular}{@{}r@{\hskip3pt}p{8cm}@{}}
        \textbf{String}: & Arbitrary string \quad \textbf{Idx}: Positive integer\\
        \textbf{Lang}: & Name of the source or target programming language\\
        \textbf{NTName}: & A valid non-terminal in the corresponding language\\
    \end{tabular}
    \caption{Translation Rule DSL for the tree transducer. }
    \label{fig:transruledsl}
\end{wrapfigure}

Formally, a transducer can be characterized by the tuple $(Q,\Sigma, \Gamma, I, \Delta)$  where $Q$ is a set of states of the transducer, $\Sigma$ and $\Gamma$ are ranked alphabets for source and target languages, $I\subset Q$ is a set of initial states, and $\Delta$ is a set of translation rules. Since our transducer is single-state there is only one state $Q=q$ and $I=q$. This tree transducer takes as input an AST tree in alphabet $\Sigma$ and generates an AST tree in alphabet $\Gamma$ using the rules specified by $\Delta$. Each rule in $\Delta$ takes the form $q(F(x_1, x_2, ..., x_n)) \rightarrow G(q(x_{i_1}), q(x_{i_2}), ..., q(x_{i_m}))$, where $F$ is a multi-level tree on $\Sigma$ with $n$ subtrees represented as $x_1, \cdots, x_n$, and $G$ is a multi-level tree on $\Gamma \cup (Q\times \{x_1,\ldots, x_n\}) $ with $m$ subtrees to be translated under state $q$, represented as $q(x_{i_k})$ and $i_k \in \{1, .., n\}$. Each $x_i$ is the placeholder matched by \textbf{Capture}, whereas $q(x_{i_j})$ is the one matched by \textbf{Reference}.

For example, consider two simple languages of simple arithmetic expressions: the source language has alphabet $\Sigma = \{\text{``}\times\text{''}, \text{``}+\text{''}, a, b, c\}$, and the target language has alphabet $\Gamma = \{\text{Add}, \text{Mult}, A, B, C\}$. Given a rule $q(\text{``}\times\text{''}(\text{``}+\text{''}(x_1, x_2), x_3)) \rightarrow \text{Add}(\text{Mult}(q(x_1), q(x_3)), \text{Mult}(q(x_2), q(x_3)))$ that matches on multiple nodes directly, in one step the transducer can translate the source tree $\text{``}\times\text{''}(\text{``}+\text{''}(a, b), c)$ in $\Sigma$ to a transduced tree in in $\Gamma$ in the form $\text{Add}(\text{Mult}(q(a), q(c)), \text{Mult}(q(b), q(c)))$. When 3 additional rules are provided as $q(a) \rightarrow A$, $q(b) \rightarrow B$ and $q(c) \rightarrow C$, then the transduction will finish and will result in $\text{Add}(\text{Mult}(A, C), \text{Mult}(B, C))$ in alphabet $\Gamma$. 
Observe that our DSL is non-deterministic as it supports learning other rules (such as $\delta_2 = q(\text{``}\times\text{''}(x_1, x_2)) \rightarrow \text{Mult}(...)$) for the same or covered LHS. Therefore, it supports ambiguous translations due to differences in the handling of language constructs and customized rules for each user based on their coding styles.

Our transducer is designed to minimize learning new rules. First, as shown in the previous example, our transducer can apply the rule over multiple nodes in the source pattern rather than just one at a time: the rule is directly applied over both nodes $\text{``}\times\text{''}$ and $\text{``}+\text{''}$ at the same time rather than applying on $\text{``}\times\text{''}$ first and then on the $\text{``}+\text{''}$. In the latter case, the final translation would have been $\text{Mult}(\text{Add}(A,B), C)$ and would have required one additional rule over the former. Second, to translate all symbols in the source alphabet $\Sigma$, every symbol would have to be addressed by some translation rule. This can be tedious if the alphabet is huge. Instead, we write (or infer) a template rule that can encompass many similar operators, identifiers, literals and so on. For example, a template node ``\texttt{\_str\_}'' (see nodes with suffix ``\textbf{Tpl}'' in Figure~\ref{fig:transruledsl}) can be created for all operators ``$<, >, \leq, \geq$'' to replace the rules corresponding to each individual operator by just one umbrella rule containing the template node.

\subsection{Syntax and Semantic Checkers}
\label{sec:validatept}

If $R$ is the set of rules that are necessary and sufficient to obtain a valid translation, then the goal is to search for a trace of rules $\tau:[r_1, r_2, \cdots, r_m], r_i \in R$ that can be applied to the source AST. This trace is conventionally called the leftmost derivation. The search for such a trace is more complicated if multiple rules can be applied for the same AST fragment which is usually the case. If there are even $2$ rules for each node in the AST then the total number of rule traces quickly raise up to $2^{\# \text{nodes}}$. Further, \tool does not know $R$ beforehand as it starts with a small set of initial rules $R_I\subset R$. Therefore, \tool's incremental procedure has to search for $R$ along with the search for the correct leftmost derivation $\tau$ to apply.

To facilitate this search, \tool uses a syntax checker to enumerate only the ASTs that lead to syntactically correct translations and uses a semantic checker to further guide the search toward finding a semantically correct translation. When \tool cannot find more rule options to apply, it prompts the user to provide code snippets to learn a new rule (later described in Section~\ref{sec:inferrule}).

\paragraph{Syntax Checker}
This component uses a pushdown automaton to keep track of the result of every transduction and reject the ones that result in syntactically invalid translations. Specifically, the automaton is parameterized by the target grammar and the transductions of every intermediate step are provided as inputs to the automaton. The automaton consumes the provided inputs and checks whether the partially translated subtree can be expressed using the target grammar. If the automaton results in an error state, then the current (latest) transduction is rejected along with all further transductions that would build on top of the current transduction.  If the automaton accepts a fully transduced AST, then a parse tree, with more information than the AST, can be reconstructed from the automaton's logs. This parse tree is pretty printed and outputted as a candidate translation.

We use parsing expression grammars (PEGs)~\cite{ford2004parsing} to define the target language syntax. A PEG can be seen as a CFG with production rule priorities. Such grammars are often easy to write and widely used in practice \cite{pep617}. With the PEG grammar and additional annotations about how the AST is mapped to the parse tree, the automaton automatically and incrementally checks an AST's syntactic validity. Formally, the non-deterministic pushdown automaton $M$ used in \tool is defined as a tuple $(Q',\Sigma',\Gamma',\Delta',q'_0,Z',F')$, where $Q'=\{\text{Start},\text{Accept},\text{Error}\}$ is a finite set of states, $q'_0 = \text{Start}$ is the start state, $F = \{\text{Accept}\}$ is the accepting state, $Z'$ is the initial stack containing the start symbol of the target PEG grammar, $\Sigma'$ is the input alphabet for representing AST as a sequence of tokens, $\Gamma'$ is the stack alphabet containing terminals and non-terminals with annotated information of those symbols, and $\Delta'$ is a set of transition instructions. The input sequence of the automaton is the preorder traversal of a partially-translated AST. The leaf nodes of the AST representing terminal symbols are allowed to be missing in the input sequence. The automaton will only enter the accept state if the stack is empty when the end of the input has been reached. We omit the details of the alphabet and transition instructions to construct such an automaton as it follows a standard textbook procedure \cite{sipser1996introduction}. 

\paragraph{Semantic Checker}
The semantic checker further helps the search for the correct translation by implementing a test-retry procedure. A simple strategy to find the correct translation among the syntactically valid candidate translations generated by the transducer would be to test all of them one by one using the unit tests. There are two issues with this strategy. First, the number of syntactically valid translations can also be potentially exponential in the code size. Second, if none of the enumerated candidate translations is correct, then the transducer might have to learn a new rule to do the right translation. But, it will only know this after testing all the enumerated candidate translations. To deal with these issues, \tool tests the candidate translations as they are being generated using the provided unit tests. When a candidate translation fails the tests, in many cases such as undeclared variables, mishandled dynamic types, and unsupported operations the location of the failure is evident. In such cases, \tool applies a new rule at that location to generate a new candidate translation. In essence, \tool sequentially repairs a portion of the code in a candidate translation and continues testing it for the next incorrect portion of the code. In practice, we observe that this strategy often finds the correct translation quickly.

It is also possible that there are no runtime errors but the outputs of the unit tests on the translated code are different from their counterparts on the source code. This kind of problem can be caused by subtle semantic differences among the built-in functions (such as \code{int(x)} vs \code{parseInt(x)} on strings), built-in operators (such as integer overflow), variable scoping (referring to the wrong variable), runtime behaviors (such as iteration order of a dictionary), and so on. In this case, if there is no line number to debug, the searcher will simply coordinate with the core to retry all possible translations until it reaches some pre-determined retry limit of the test-retry loop or returns the first translation that passes unit tests. As future work, other fault localization techniques~\cite{artzi2010practical,ocariza2016automatic,shen2021localizing} could improve our approach for such cases.
\begin{todoenv}
\paragraph{Search Optimization}
To make the search and check more efficient, we use lazily constructed transduction trees, dynamic programming, and caching of states of the pushdown automaton for efficient backtracking. With such optimization, the execution time to search for new candidate translations is negligible compared to the running time of the test-retry procedure. The precise details of the optimization are provided in the supplementary material. \end{todoenv}

\subsection{Inferring Translation Rules}
\label{sec:inferrule}

\begin{algorithm}[t]
    \caption{Rule Inference Algorithm}
    \label{alg:ruleinfer}
    \begin{algorithmic}[1]
      \Procedure{RuleInfer}{SrcASTFragments, TarASTFragments}
      \State LHS = \Call{SimultaneousTraversal}{SrcASTFragments} \Comment{AST template for source} \label{alg:line:srctraversal}
      \State src\_PHs, src\_tpl\_PHs = \Call{getPlaceholders}{LHS}\label{alg:line:lhsplaceholder}
      \State RHS = \Call{SimultaneousTraversal}{TarASTFragments}\Comment{AST template for target}\label{alg:line:tartraversal}
      \State tar\_PHs, tar\_tpl\_PHs = \Call{getPlaceholders}{RHS}\label{alg:line:rhsplaceholder}
      \For {tar\_PH \textbf{in} tar\_PHs} \Comment{Compute pairwise tree-edit distance source-target}
        \State refer\_idx = $\mathop{\text{argmin}}_{i}$(\Call{TreeEditDistance}{src\_PHs[i], tar\_PH})
        \State tar\_PH.\Call{setReferenceIdx}{refer\_idx} \Comment{Find the smallest distance for each target PH}
      \EndFor
      \For {tar\_tpl\_PH \textbf{in} tar\_tpl\_PHs} \Comment{Match terminals}
        \State tpl\_idx = $\mathop{\text{argmin}}_{i}$((src\_tpl\_PHs[i] == tar\_tpl\_PH) ? 0 : 1)
       \State tar\_PH.\Call{setTemplateIdx}{tpl\_idx == None ? ``Unexpected'' : tpl\_idx}
      \EndFor
      \State \Return \Call{createRule}{LHS, RHS} \Comment{Wrap LHS, RHS in the DSL format rule}
      \EndProcedure
    \end{algorithmic}
\end{algorithm}

After the steps of the syntactic and semantic check, when \tool does not find a suitable rule to apply, it proceeds to infer a new rule. 
The rule inference problem consists of producing a pair of syntactic patterns and a corresponding translation rule from the user-provided code pairs of source-target language snippets.
\tool uses one or at most two pairs of snippets. 
The inferred rule has the source pattern (AST template) on the left-hand side and a target pattern (AST template) on the right-hand side, as shown in Figure~\ref{fig:ruleinfer}. 

The rule inference problem consists of  three sub-problems. 
The first problem is to extract the ASTs for the instances of the pattern that appear in both the source and target code snippets. The second problem is to extract one AST template that summarizes both the ASTs in each code snippet. The third problem is to generate a rule that translates the AST template from the source snippets to its corresponding one from the target snippets.

\paragraph{From Code Snippets to ASTs.} To address the first problem, \tool assumes that a user provides only one instance capturing the syntactic pattern of interest per line and any additional code not matching that syntactic pattern is minimal. 
Further, it assumes that all of the instances in both source and target snippets are either expressions or statements as it is unlikely to translate between expressions and statements. 
\tool then automatically extracts the AST for each instance, representing an expression or a statement, for every code snippet. 
For example, in Figure~\ref{fig:ruleinfer}, the ASTs are automatically extracted from \pylang and \jslang expressions from each line. If the extraction fails due to identifying unintended instances, 
\tool also allows users to highlight the intended instances that capture the syntactic patterns.

\paragraph{Extracting AST Templates} Given the resulting ASTs from source and target code snippets, our rule inference procedure infers a translation rule. The algorithm for rule inference is given in Algorithm~\ref{alg:ruleinfer}. To find the AST template for the ASTs in the source code snippets the algorithm finds the largest common AST fragment between the source ASTs. Specifically, our algorithm at line~\ref{alg:line:srctraversal} calls  the \textsc{SimultaneousTraversal} subprocedure\footnote{Refer to the supplementary material for details.} which matches the nodes of the two ASTs by using simultaneous pre-order traversals on both of them. The subtrees that are different will be abstracted away and replaced by placeholders (or template placeholders if the differing subtree is a terminal) in the extracted AST template (see line~\ref{alg:line:lhsplaceholder}). The source AST template is the LHS of the rule being inferred (see left of the translation rule in Figure~\ref{fig:ruleinfer}). After that, the same \textsc{SimultaneousTraversal} procedure is applied to target ASTs (line~\ref{alg:line:tartraversal}) to find the AST template for the RHS of the rule. 

\paragraph{AST Templates to Rules.}
Recall that according to our rule DSL (see Section~\ref{sec:transpacerule}) the placeholders on the LHS are \textbf{Capture}s and the placeholders in the RHS are \textbf{Reference}s. The \textbf{Reference}s and template expanders in the RHS are still placeholders and have not been connected to \textbf{Capture}s and template matchers in the LHS yet. 
For each template expander, the connection to template matchers is simply computed by string comparisons (see lines $9$-$11$). We next connect the \textbf{Reference}s with \textbf{Capture}s in the template AST (see lines $6$-$8$). The idea is to connect a \textbf{Reference} to the \textbf{Capture} that is most similar to it based on the subtrees they encode. Each \textbf{Reference} encodes two subtrees, one each from the target ASTs and each \textbf{Capture} encodes two subtrees, one each from the source ASTs. The similarity between a \textbf{Capture} and a \textbf{Reference} is computed by summing up the pairwise similarities between their corresponding subtrees. We use tree-edit distance as a similarity metric to compute the similarity between two subtrees. The tree-edit distance is affected by the subtree structure, node names, and node values in the terminals. The tree edit distance is computed using the APTED algorithm~\cite{pawlik2016tree} and the distance between nodes is defined as Levenshtein string edit distance for node names and terminal node values~\cite{levenshtein1966binary}. The obtained rule for our running example is given in Figure~\ref{fig:ruleinfer}.
The connections between \textbf{Reference}s and \textbf{Capture}s are inferred based on an assumption that the translation of a specific subtree in the source AST has the smallest tree edit distance to its corresponding subtree in the translated AST compared to all other translated subtrees. In practice, we find that this assumption holds for almost all of the code snippets we provided. When it does not hold, \todoa{the user may manually choose corresponding \textbf{Reference}s of \textbf{Capture}s in a rule editor.}

\todoa{To summarize, it is up to the user to decide if the rule-inference procedure is automatic or manually guided. 
Furthermore, 
the rule-inference procedure is not particularly specific to Python and JavaScript, and in a similar fashion can work with other languages as all it needs is the ability to construct ASTs out of code snippets provided in the two languages.}

\section{Implementation}
\label{sec:tool}

 \tool is open source \cite{supp}. It is implemented in a modular fashion. \tool's core module implements the tree transducer and the \syntaxchecker \todoa{with optimizations mentioned in supplementary materials.}
The core module exposes an interface for interacting with external procedures to select rules during the search for correct translation. \tool also contains peripherals that enable automating the test-driven incremental rule learning. Furthermore, it supports a plug-and-play user interface that allows users to control and customize all phases of translation.

\subsection{\toolbf Core}
\label{sec:toolcoreimpl}

\begin{figure}[ht]
    \centering
    \includegraphics[width=0.7\linewidth]{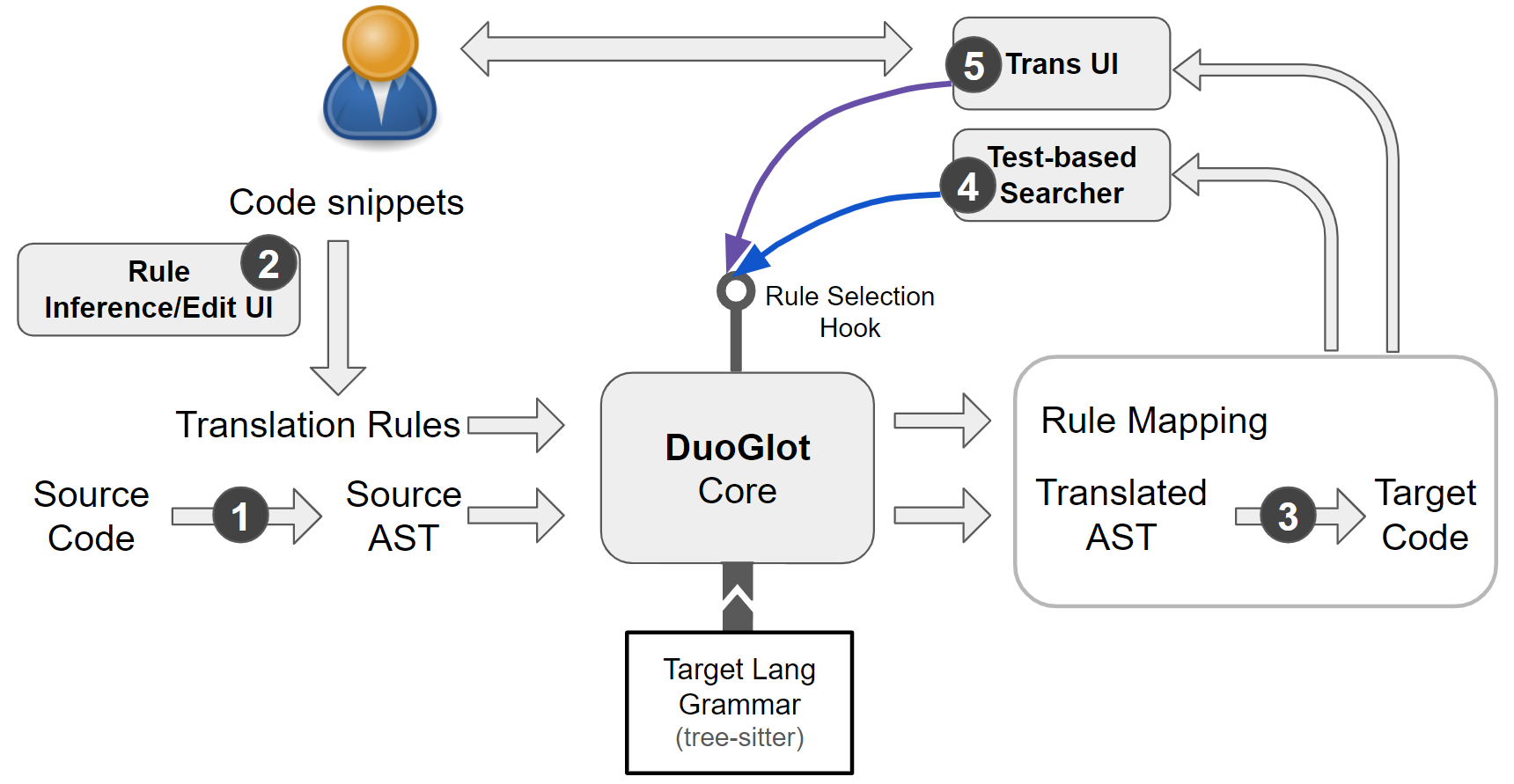}
    \caption{\tool's modules and inputs/outputs.}
    \label{fig:duoimpl}
\end{figure}

\textbf{Input and Output}. \tool's core focuses on translating the AST of the source code to an AST in the target language. It requires two inputs: the AST of the source code and a set of rules to start the translation procedure (see Figure~\ref{fig:duoimpl}). The AST for source code is generated by a Tree-sitter parser~\footnote{\href{https://tree-sitter.github.io/tree-sitter/}{Tree-sitter} is a popular parser-generator used in production. It accepts PEG grammar definition.} for the source language. \tool also requires the PEG grammar definition for the target language which can be obtained from the Tree-sitter package for the target language. \tool finds a trace of rules that transduce the input AST to a syntactically-valid AST in the target language. It outputs such a candidate AST along with a mapping of rules that have been used to generate it. 

\noindent\textbf{Rule Selection Hook}. The core implements a non-deterministic tree transducer so there might be multiple rules that are applicable at the same translation step. Some applications are wrong or may not be the preferred ones, depending on the source program, user's intention and the unit tests. Therefore, the hook allows for arbitrary external modules or search heuristics to interact with \tool's core to select or prioritize rules that are applicable at the same translation step. The user can also choose the rules that can lead to an intended translation.

\noindent\textbf{Parametrization}. The \syntaxchecker inside the core is parameterized by a parser expression grammar (PEG) definition for the target language. We implement the checker to accept Tree-sitter's grammar DSL for that purpose. The tree-sitter grammar definitions for many common programming languages are already used in production \cite{linktsbloga}. Therefore, \tool's core can be easily extended to support other scripting languages with existing Tree-sitter grammar definitions.

\subsection{\tool Peripherals}
\label{sec:peripherals}

\tool's core outputs a candidate syntactically-valid AST and does not contain the test-repair procedure or the modules to convert between AST and exact code. These features are implemented in the peripheral modules (see Figure~\ref{fig:duoimpl}). 

\noindent\textbf{Parser and Pretty Printer}. Circles 1 and 3 in Figure~\ref{fig:duoimpl} represent the parser for the source language and pretty-printer for the target language respectively. For the parser, we use off-the-shelf Tree-sitter parser to get AST from source code. The pretty-printer is used to convert the translated AST, in the Tree-sitter format, to code in the target language. There are no existing implementations for such pretty-printers. So we implement a meta pretty-printer that is parameterized by the Tree-sitter's PEG for the target language. In addition, we also handle some corner cases for each language such as terminals and non-context-free special cases (such as \pylang's indentation).

\noindent\textbf{Test-based Searcher}. When the \tool's core generates a candidate translation it is passed on to the test-based searcher. The searcher runs the unit tests to verify its correctness. Circle 4 in Figure~\ref{fig:duoimpl} shows the test-based searcher implementing the testing of the candidate translations using the unit tests and error analysis.
If the translation passes all the unit tests, the searcher will output the translation. If the translation fails unit tests with a fault line number, the searcher will guide its search in the following way. It will first look at the rule mapping to compute the segment of the derivation history that corresponds to the faulty line of code.  Then the searcher will choose other possible rule selections in that segment and send the choices back to the \tool core. Note that there could be multiple rule applications corresponding to one line of code. Therefore, in the worst case, the searcher will have to try all possible combinations of applicable rules on that line. Hence, the searcher will first choose to change only one rule at a time,  then proceed to change two rules and so on.  When the core receives the updated choices, it will produce another candidate translated AST that obeys the new rules selected by the searcher. The new candidate translation will be tested again. If the error on that line cannot be fixed after trying all alternatives on the fault line, the searcher will get stuck and prompt the user for new code snippets. 

We impose a limit on number of test-retry iterations, called the retry limit. For instance, in our main evaluation we set the retry limit to $20$. So within these, say $20$, retries, 1) the \tool may produce a translation that passes all unit tests, 2) it may get stuck because it has exhausted all rules to apply at a line, 3) all the $20$ candidate translations raise runtime errors, or 4) the last candidate translation does not have runtime errors but still fails to have the same outputs as the source code. Only when the \tool gets stuck (second scenario), the user is prompted for new code snippets to learn a new rule and it restarts its test-retry procedure again.

\noindent\textbf{Trans UI (User-based Rule Selection)}.  Circle 5 in Figure~\ref{fig:duoimpl} represents a component of \tool's user interface. The component allows users to see the translation procedure. Users can inspect a candidate translation and its rule mapping. Users can also change the rules at each translation step during the translation procedure to obtain a different candidate translation. A screenshot of this component is given in  the supplementary material.

\noindent\textbf{Rule Inference/Edit UI}. Circle 2 in Figure~\ref{fig:duoimpl} represents the rule inference/edit component of \tool's UI. Here, users input the code snippets and inspect the automatically inferred rules. Expert users may also manually edit the rules if they want. A screenshot of this component is provided in the supplementary material.

\section{Evaluation}
\label{sec:eval}

We show that \tool is one of the best practical tools for translating between untyped scripting languages such as Python and JavaScript. 
We plan to submit an artifact with our code and benchmarks.
Our evaluation is structured as follows.

\begin{enumerate}
    \item \textbf{Quantitative (Sections~\ref{sec:eval-tool-perf} and ~\ref{sec:eval-tools-comp})}: We measure the number of benchmarks that \tool can solve correctly and the number of rules it requires to solve them. We then compare \tool to several existing translators on correct translations and their readability.
    \item \textbf{Qualitative (Section~\ref{sec:eval-human-effort})}: We manually categorize the rules based on their complexity and identify how the rule complexity affects performance. We discuss as well the additional human effort required by \tool as compared to other state-of-the-art transpilers.
    \item \todoa{\textbf{Case Studies (Section~\ref{sec:caseleet} and \ref{sec:casectci})}: 
    We present two case studies to evaluate the abilities and limitations of our tool in scenarios where the test coverage varies and the code is either more complex or longer. 
}
\end{enumerate}

\subsection{Evaluation Setup}
\label{sec:eval-setup}

We evaluate \tool for translating code from \pylang to \jslang on popular benchmarks. 

\noindent \textbf{Benchmarks.} 
\todoa{We use GeeksForGeeks (GFG) benchmarks ~\cite{linkgfg} from the Transcoder project~\cite{roziere2020unsupervised} as our main evaluation benchmarks, and 
two additional sets of programs for our case studies
(refer to Section \ref{sec:caseleet} and \ref{sec:casectci}).} The GFG benchmarks consists of $702$ standalone \pylang scripts with unit tests, each having a  target function for translation. We remove $3$ scripts that we could not run \todoa{(contain errors)}, resulting in $699$ \pylang scripts to evaluate on. 
The target functions vary in lengths in the range of $1$-$40$ lines ($8.53$ on average), or $23$-$974$ non-space characters ($154$ on average). 

\noindent \textbf{Unit tests.} 
The benchmarks provide unit tests for the source \pylang code \todoa{(which achieve 99\% line coverage)}, but not for the target \jslang. 
\todoa{However, \pylang unit tests are straightforward to translate to \jslang, 
either with a simple ad-hoc translator for unit tests as done in previous works~\cite{roziere2020unsupervised} or with a simplified version of \tool (without test-repair strategy). We use the latter and manually verify that all of the unit tests have been correctly translated.  }

\noindent \toolbf \textbf{Setup.} 
To translate \pylang to \jslang, we create a pipeline 
using \tool Core and \tool Peripherals described in the previous section. We turn off the part of the user interface that allows manually choosing alternative rules and set the retry limit for the test-retry loop (running unit tests) to  $20$ iterations. We choose $20$ as we want \tool  either to finish translating the source code within $1$-$2$ seconds, or if it fails, to ask for new code snippets.
At the beginning, \tool starts with \initialsetcount base rules for \todoa{direct mapping of the most common AST nodes, such as identifiers, number literals, string literals, assignments, etc. from \pylang to \jslang}. \todoa{We include these base rules by default (even though end users can also create or modify them), to avoid repeated work of redefining them in all current and future applications of \tool.
}

\noindent\textbf{Library Calls. } \todoa{
The GFG benchmarks depend only on built-in Python libraries and do not make calls to third-party libraries. 
Limiting calls only to built-in libraries 
is common when testing Python to JavaScript translations as it allows effective comparison of core capabilities of transpilers. Note, \tool does not have fundamental limitations in handling third-party libraries (such as NumPy and pandas), however, it will require the existence of similar libraries in JavaScript.}

\noindent \textbf{Evaluating accuracy.}
To confirm the correctness of the translations we use unit tests. We compare the outputs of the source and the translated programs on unit tests, including their intermediate logs (prints).  We say a translation is correct only if the outputs match on all unit tests. Accuracy then is the percentage of benchmarks that have been correctly translated. 

\noindent \textbf{Evaluating readability.} We assess the readability based on the translated code size~\cite{weinberg1971psychology} \todoa{and assume smaller size is better because it preserves high-level abstractions as it precludes translations that have many additional unnecessary variables or redundant code. 
We prefer this to other readability metrics (such as~\cite{buse2008metric}), because our main interest is to stay at the same abstraction level as the source program, rather than to focus on human factors such as identifier naming, quality of comments, etc.}
We express the readability as the bloating ratio between translated code and the source code of the number of non-whitespace characters.

\noindent \textbf{Compared transpilers.} We compare \tool with other state-of-the-art transpilers, both ML-based and hand-crafted. Among ML-based transpilers, we evaluate Codex\cite{chen2021evaluating} (the API used by Github Copilot) and TransCoder\cite{roziere2020unsupervised}. For hand-crafted transpilers, we search for open source projects on GitHub that are relevant to \pylang to \jslang translation. 
We choose Transcrypt ($2.5$k stars\footnote{Number of stars means number of developers on GitHub that keep track of the project.}), Javascripthon\footnote{Python 3 to ES6 Javascript translator. \url{https://github.com/metapensiero/metapensiero.pj}} ($824$ stars), and py2js\footnote{Python to JavaScript translator. \url{https://github.com/qsnake/py2js}} ($90$ stars) as they also aim to produce correct and readable \jslang code.
To avoid unfair comparison, we use the following criteria that generally \emph{overestimate} the performance of some of these tools:
 \begin{itemize}
\item \textbf{TransCoder}: 
     We take the $10$ most probable translations (i.e., with beam size $10$) and run tests on all $10$ translations. We count it as successful if any of the $10$ translations can pass all the unit tests.
\item \textbf{Codex}: We take one possible translation from Codex API. We clean up the translation by removing the additional code or text after the first \jslang function in the translation. If there is a typo in the translated function name, we fix it by string replacement.  We count it as successful if the cleaned and fixed translation can pass all the unit tests.
\item \textbf{Transcrypt, Javascripthon, and py2js}: We run them once for each benchmark to get the translated code.  We count it as successful if the translation passes all the unit tests.
 \end{itemize}

\noindent \textbf{System Specification}. We use a desktop with an Intel i7-9700 8-core CPU and 32GB of RAM. The \tool Core, test-based searcher, and the evaluation script for the non-ML-based tools are run on one CPU core each. 
For TransCoder, we use a machine with AMD Threadripper 3970X 32-Core Processor with 64GB of RAM and a GeForce RTX 3090 GPU. For Codex, we use OpenAI's API directly without any local computation.

\subsection{Performance of \toolbf \todoa{on GFG benchmarks}} 
\label{sec:eval-tool-perf}

\begin{figure}[ht]
    \centering

\begin{minipage}{.45\textwidth}
  \centering
  \includegraphics[height=90pt]{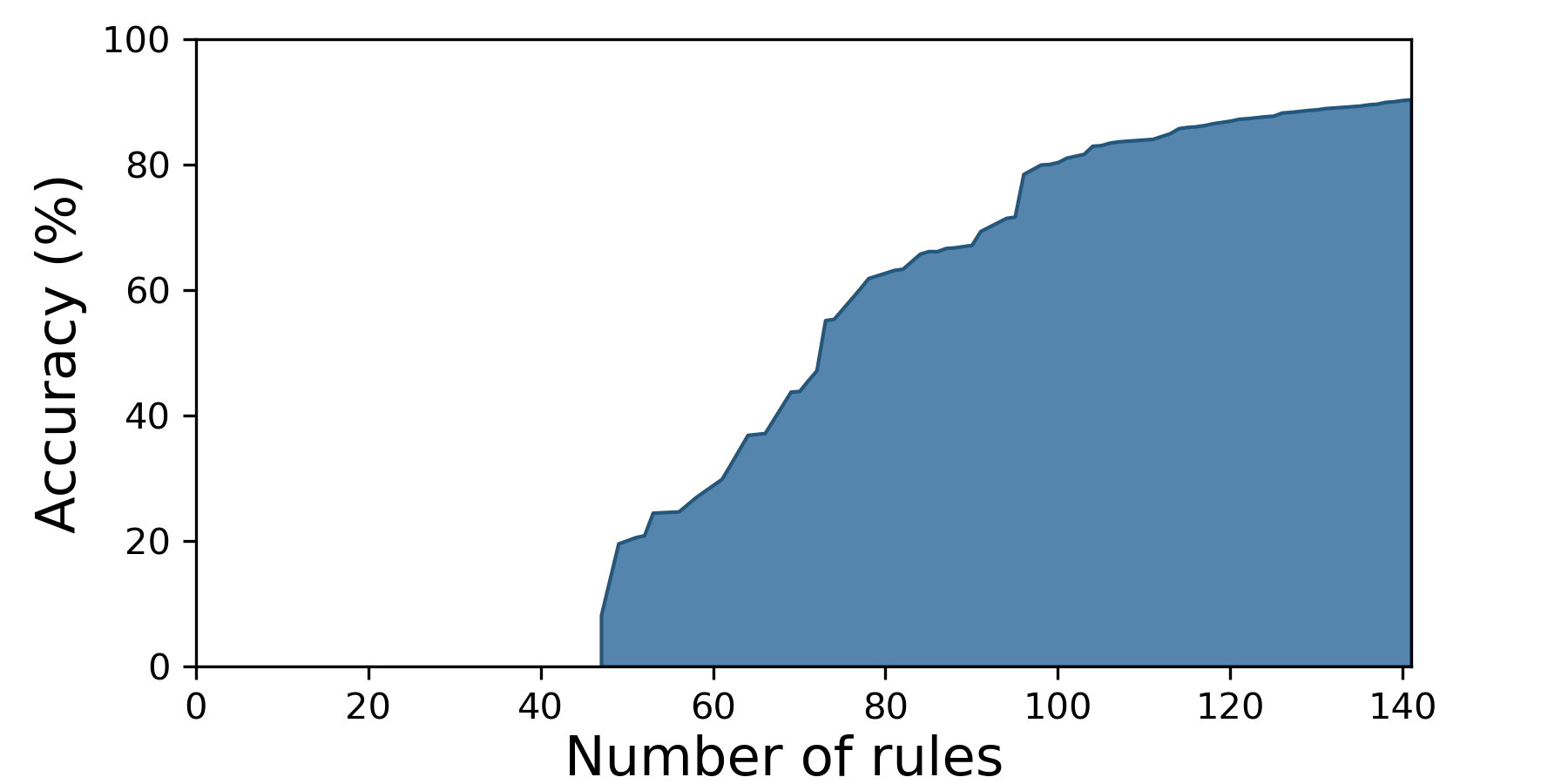}
  \caption{Accuracy of the translated benchmarks from \gfgbenchmark as a function of the number of used rules. For the first $80$ rules, the accuracy increases from $0\%$ to $60\%$. }
  \label{fig:eval-accuracy}
\end{minipage}%
\hfill
\begin{minipage}{.5\textwidth}
  \centering
  \includegraphics[height=90pt]{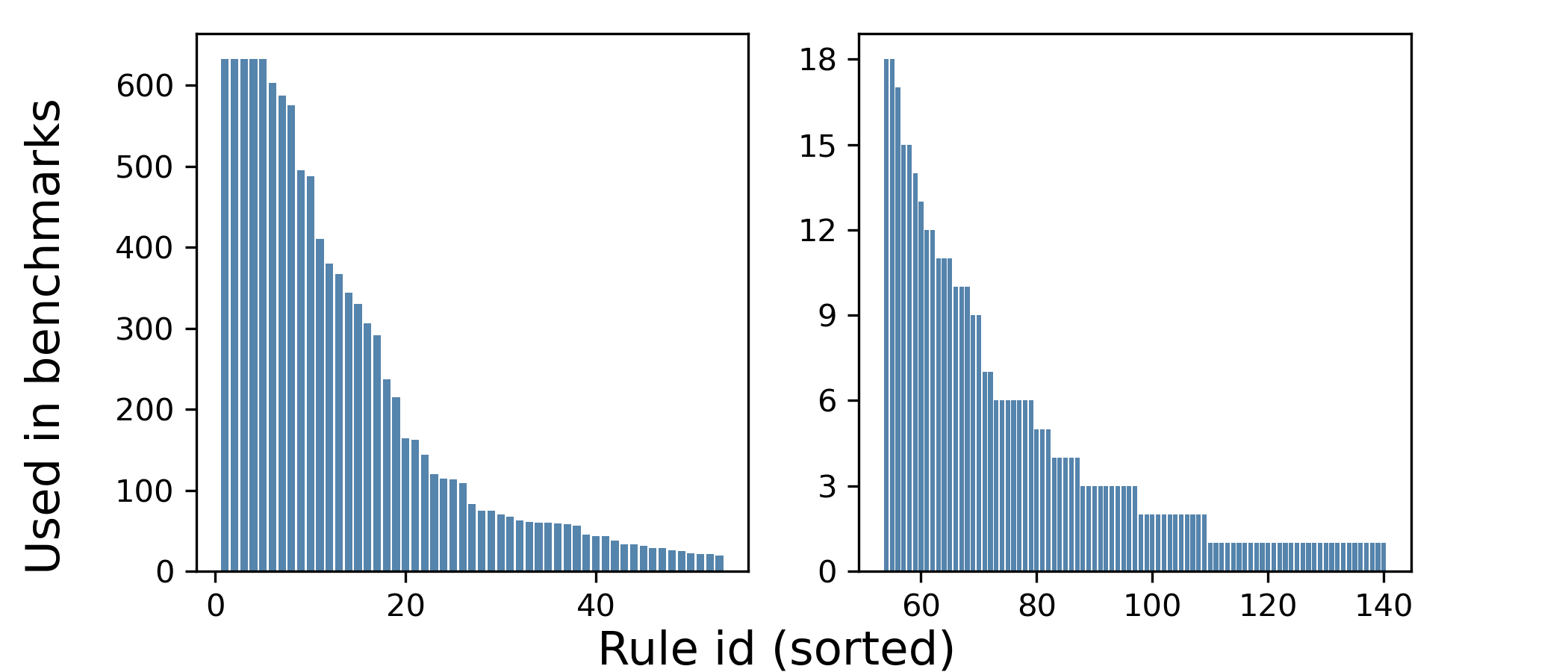}
  \caption{Distribution of the \totalrulecount rules across the translated benchmarks. The top 15 most popular rules are presented in around half of the benchmarks, and all of these rules are the base rules. }
  \label{fig:eval-rules}
 \end{minipage}
 \end{figure}

\tool increases its accuracy by inferring rules learned from user-provided code snippets. With no learned rules (only the bootstrapped \initialsetcount rules mentioned in the setup), its accuracy is $0$\% on the $699$ benchmarks, but this grows to $90$\% with \totalrulecount total rules. 
We plot the relation between the number of rules and the percentage of benchmarks that can be solved by  a specific number of rules in Figure~\ref{fig:eval-accuracy}. %~\ref{fig:rulecountperf}. 
As evident, the accuracy quickly increases from $0$\% to around $60$\% as the number of total rules reaches $80$, and $80$\% with $98$ rules. After that, the accuracy increases slowly and it reaches the top $90$\% at \totalrulecount rules. 

We provide as well the distribution of the rules across the translated benchmarks, refer to Figure~\ref{fig:eval-rules}. %~\ref{fig:rulecountperf}. 
The top $15$ most popular rules are present in up to one-half of the benchmarks. As expected, all of these rules are from the bootstrap set (base rules). The usage of the remaining rules varies, from hundreds of benchmarks for the more used, to only a handful for the less applied.

\subsection{Comparison to Other Transpilers \todoa{on GFG benchmarks}}
\label{sec:eval-tools-comp}

Further, we provide comparisons of the features of all transpilers and refer the reader to Table~\ref{tab:testacc}.

\noindent \textbf{Accuracy Comparison}. The accuracy of all the transpilers under the considered evaluation setting is shown in Table~ \ref{tab:testacc}. \tool achieves better performance than all hand-crafted transpilers such as Transcrypt, Javascripthon, and py2js. Furthermore, it also outperforms Codex even  after fixing minor typos in the translations. For TransCoder, a direct comparison is not possible, so for reference we only list TransCoder's performance from Python either to C++ or to Java. 
Due to the lack of data and resources to re-train TransCoder for \jslang target, we can only speculate that when trained on \jslang, TransCoder's performance on \pylang-\jslang translations will be comparable to that of its \pylang-C++ translation.

\begin{table}[ht]
\caption{Comparison of accuracy, readability (bloating ratio) and execution times of different transpilers. } 
\label{tab:testacc}
\centering
\small
\begin{tabular}{|l|c|c|c|}
\hline
\textbf{Translate Python to Javascript} & \textbf{Accuracy} & \textbf{Readability} & \textbf{Time} \\
\hline
\hline
\tool (\totalrulecount rules) & 90\% & 1.34x & 0.46s  \\
Codex \footnotemark & \codexrawacc & 1.2x & 7.33s  \\
Codex (fixed typos) \footnotemark  & \codexfixacc & 1.2x & 7.33s \\
Transcrypt & \tscryptacc  & 10.7x\footnotemark & 0.16s \\
Javascripthon & \jthonacc & 2.1x & 0.06s \\
py2js & \pyjsacc & 2.2x & 0.03s \\
\hline
\hline
\textbf{Other Settings} & \textbf{Accuracy}  & \textbf{Readability} & \textbf{Time} \\
\hline
\hline
TransCoder (to Java, beam size 10) & 67\% & - & \transcoderjavatime s  \\
TransCoder (to C++, beam size 10) & 79\% & - & \transcodercpptime s \\
\hline
\end{tabular}
\end{table}

\addtocounter{footnote}{-1}
\addtocounter{footnote}{-1}
\addtocounter{footnote}{-1}
\stepcounter{footnote}
\footnotetext{The model we are using is \texttt{code-davinci-001} and the prompt we are using is \texttt{'\#\#\#\#\# Translate this function from Python into JavaScript\textbackslash n\#\#\# Python\textbackslash n\textbackslash n<Python Code>\textbackslash n    \textbackslash n\#\#\# JavaScript\textbackslash n"use strict";\textbackslash n'}}
\stepcounter{footnote}
\footnotetext{For many of the benchmarks translated by Codex, the function name is wrong and we fix it by string replacement.}
\stepcounter{footnote}
\footnotetext{Transcrypt prepends a lot of ``imports'' in its translations for emulating Python functions in JavaScript.}

\noindent \textbf{Readability Comparison}. 
\tool provides as well one of the top readabilities (quantified as the bloating ratio) among all the tested transpilers (refer to Table~\ref{tab:testacc}). On average across all accurately translated benchmarks, \tool increases the size of the original \pylang program into the corresponding \jslang program by a factor of $1.34\times$ (varies in the range $1.06$-$2.22$).  
That is best only next to Codex's $1.2\times$ readability. On the other hand, the remaining transpilers have a substantially larger bloating ratio:  $10.7\times$ for Transcrypt without considering the size of the API emulation library, $2.1\times$ for JavaScripthon, and $2.2\times$ for py2js.

\noindent \textbf{Time Comparison}. Once \tool obtains all the rules to translate a benchmark, 
the total average time spent to translate a benchmark including testing is $0.46$s.
The average time for the other manual transpilers is as follows: py2js requires 0.03s, JavaScripthon $0.06$s, and Transcrypt $0.16$s. The time for TransCoder to perform \pylang to C++/Java translation, with beam size $10$, is $\transcodercpptime$/$\transcoderjavatime$s on average per benchmark. Finally, API calls to OpenAI's Codex require $7.43$s on average per benchmark, including the round trip time. Thus the execution time of \tool is between hand-crafted and ML-based transpilers. Note, the former translates faster due to the lack of a test-repair loop. This loop in \tool, however, provides a time-accuracy tradeoff: at the expense of extra computations, significantly improves the correctness of the translation. As such, it is arguably beneficial as the execution time is still very practical.

\subsection{Rules Classification and Manual Effort}
\label{sec:eval-human-effort}

\begin{myverbbox}[\scriptsize]{\initialverbpya}

a = 5
[c, d] = [e, f]
\end{myverbbox}

\begin{myverbbox}[\scriptsize]{\initialverbjsa}

a = 5;
[c, d] = [e, f];
\end{myverbbox}

\begin{myverbbox}[\scriptsize]{\initialverbpyb}

[]
[1,2,3]
\end{myverbbox}

\begin{myverbbox}[\scriptsize]{\initialverbjsb}

[];
[1,2,3];
\end{myverbbox}

\begin{myverbbox}[\scriptsize]{\initialverbpyc}

None
\end{myverbbox}

\begin{myverbbox}[\scriptsize]{\initialverbjsc}

null;
\end{myverbbox}

\begin{myverbbox}[\scriptsize]{\simpleverbpya}

not x
not y == 0
\end{myverbbox}

\begin{myverbbox}[\scriptsize]{\simpleverbjsa}

!(x);
!(y === 0);
\end{myverbbox}

\begin{myverbbox}[\scriptsize]{\simpleverbpyb}

"hello".isupper()
s.isupper()
\end{myverbbox}

\begin{myverbbox}[\scriptsize]{\simpleverbjsb}

"hello" === "hello".toUpperCase();
s === s.toUpperCase();
\end{myverbbox}

\begin{myverbbox}[\scriptsize]{\simpleverbpyc}

a in xs
2 in []
\end{myverbbox}

\begin{myverbbox}[\scriptsize]{\simpleverbjsc}

xs.indexOf(a) >= 0;
[].indexOf(2) >= 0;
\end{myverbbox}

\begin{myverbbox}[\scriptsize]{\mediumverbpyb}

list1.remove("bob")
[1,2,3].remove(1)
\end{myverbbox}

\begin{myverbbox}[\scriptsize]{\mediumverbjsb}

list1.splice(list1.indexOf("bob"), 1);
[1,2,3].splice([1,2,3].indexOf(1), 1);
\end{myverbbox}

\begin{myverbbox}[\scriptsize]{\mediumverbpyc}

for x in xs: pass
for y in [1,2,3]: 
  break
\end{myverbbox}

\begin{myverbbox}[\scriptsize]{\mediumverbjsc}

for(let x of xs) {}
for(let y of [1,2,3]) {break;}
\end{myverbbox}

\begin{myverbbox}[\scriptsize]{\mediumverbpyd}

for x in range(10): pass
for y in range(a): 
  break
\end{myverbbox}

\begin{myverbbox}[\scriptsize]{\mediumverbjsd}

for (let x = 0; x < 10; x++) {}
for (let y = 0; y < a; y++) {break;}
\end{myverbbox}

\begin{myverbbox}[\scriptsize]{\complexverbpy}

{x["id"]:x for x in xs}
{y:y*y for y in [1,2,3]}
\end{myverbbox}

\begin{myverbbox}[\scriptsize]{\complexverbjs}

Array.from(xs).map((x) => [x["id"], x])
  .reduce((a, b) => (a[b[0]] = b[1], a), {});
Array.from([1,2,3]).map((y) => [y, y*y])
  .reduce((a, b) => (a[b[0]] = b[1], a), {});
\end{myverbbox}

\begin{table}[!htbp]
    \caption{Examples of snippets for rules from \textsf{Base}, \textsf{Simple}, \textsf{Medium}, and \textsf{Complex} categories}
    \label{tbl:rulecategory}
    \centering
    \begin{tabular}{|@{}l@{}|@{}l@{}|@{}l@{}|}
    \hline 
    \mycentered{Category} & \mycentered{\pylang code snippets} & \mycentered{\jslang code snippets } \\
    \hline 
    \hline 
    \mycentered{Base} &
    \mycentered{\initialverbpyc} & \mycenteredbuf{\initialverbjsc}
    \\
    \hline 
    \mycentered{Base} &
    \mycentered{\initialverbpya} & \mycentered{\initialverbjsa}
    \\
    \hline 
    \hline
    \mycentered{Simple} &
    \mycentered{\simpleverbpya} & \mycentered{\simpleverbjsa}\\
    \hline 
    \mycentered{Simple} &
    \mycentered{\simpleverbpyc} & \mycentered{\simpleverbjsc}\\
    \hline 
    \hline
    \mycentered{Medium} &
    \mycentered{\mediumverbpyb} & \mycentered{\mediumverbjsb}\\
    \hline 
    \mycentered{Medium} &
    \mycentered{\mediumverbpyd} & \mycentered{\mediumverbjsd}\\
    \hline 
    \hline
    \mycentered{Complex} &
    \mycentered{\complexverbpy} & \mycentered{\complexverbjs}\\
    \hline 
    \end{tabular}
\end{table}

\tool achieves $90\%$ accuracy on the tested benchmarks with \tot rules. 
Aside from the \boot base rules, we can categorize the remaining $98$ rules based on their complexity into three classes: \textsf{Simple}, \textsf{Medium} and \textsf{Complex}. The complexity is determined by the number of occurrences of \textbf{Capture}s and \textbf{Reference}s (refer to Section~\ref{sec:approach}):  the higher the number of occurrences, the more complex code snippets (in terms of length) are required to infer the rule. \textsf{Simple} rules have $2$-$6$ occurrences, \textsf{Medium} $7$-$9$, while \textsf{Complex} rules have more than $9$ occurrences of \textbf{Capture/Reference}.  %
We emphasize that the complexity of the rules does not affect the number of needed code snippets for their inference, which is always one or two.
In total, among the 98 rules,
\rulecountsimple ($84\%$) are \textsf{Simple} rules, \rulecountmedium ($9\%$) are \textsf{Medium}, and \rulecountcomplex ($7\%$) are \textsf{Complex} rules. 
Hence, among all rules, an overwhelming majority are simple. In Table~\ref{tbl:rulecategory}, we provide examples of rules from each category, described with their inferring code snippets.

The manual effort required to run \tool can be divided into two types. The first type dominates the effort and consists of providing the aforementioned code snippets used for inference of new rules for the incremental transpiler.  
Across all rules, we provided in total only \snippetlc lines of snippets (for \emph{both} \pylang and \jslang) containing \snippetcc non-empty characters. 
The second type of effort is the occasional manual adjustment of the rule inference, when the expected rule differs from the inferred. We performed this adjustment for $15$ rules. Each adjustment consists of a handful of clicks (depending on the rule complexity) in the rule editor to identify the connections between \textbf{Reference}s and \textbf{Capture}s, as specified in Section~\ref{sec:inferrule}. The second type of effort can be eliminated completely, with more advanced implementation techniques. 
Finally, we emphasize that the manual effort is a one-time cost per rule, and distinct rule sets can be merged and expanded.

\begin{todoenv}
\subsection{Case Study: Translating Solutions for LeetCode Programming Exercises}
\label{sec:caseleet}

\begin{table}[ht]
\caption{\tool's Performance in Case Studies } 
\label{tab:caseleet}
\centering
\footnotesize
\begin{tabular}{cc}

    \begin{tabular}{|l|l|c|}
    \hline
    \textbf{Case Study: LeetCode} & \textbf{Accuracy\footnotemark} & \textbf{Time}  \\
    \hline
    \hline
    \tool (142 + 5 rules) & 37.2\% / - & -  \\
    \tool (142 + 16 rules) & \textbf{51.7\%} / - & -  \\
    \tool (142 + 41 rules) & \textbf{61.4\%} / - & -  \\
    \tool (142 + 106 rules) & \textbf{75.4\% / 71.6$\pm$5\%} & 1.75s  \\
    Codex (fixed typos)  & 50.0\% / - & 5.23s \\
    Transcrypt & 48.7\% / -  & \textbf{0.24s} \\
    Javascripthon & 0\% / - & - \\
    py2js & 0\% / - & - \\
    \hline
    \end{tabular}

& 

    \begin{tabular}{|l|c|c|}
    \hline
    \textbf{Case Study: CtCI} & \textbf{Accuracy} & \textbf{Time}  \\
    \hline
    \hline
    \tool (142 + 34 rules) & 28\% & -  \\
    \tool (142 + 55 rules) & \textbf{56\%} & 6.78s  \\
    Codex (fixed typos)  & 12\% & 37.6s \\
    Transcrypt & 44\%  & \textbf{0.25s} \\
    Javascripthon & 8\% & - \\
    py2js & 0\% & - \\
    \hline
    \end{tabular}

\end{tabular}
\end{table}

\footnotetext{Offline accuracy / online accuracy. Offline accuracy is the test accuracy on example tests. Online accuracy is computed as 98\% confidence interval from a sample of 100 programs manually submitted to LeetCode.}

LeetCode\footnote{\url{https://leetcode.com/problemset/all/}, \url{https://www.linkedin.com/company/leet-code}} is a popular, multi-million user online platform for coding contests and problems. There are more than 2000 problems, to which users can submit solutions in several supported programming languages. 
Leetcode checks the solutions on correctness internally with multiple unit tests, while allowing public access only to a handful of unit tests per problem. 
Users usually share their solutions on other platforms such as Github, in particular  \url{https://github.com/doocs/leetcode} has 15k+ stars.
Most of these user-provided solutions are written in Python, but fewer in JavaScript. 
Hence, programmers in JavaScript may be at a disadvantage as they cannot fully benefit from the  numerous Python solutions.  

We use \tool to make available the solutions  in JavaScript.  
We take 1414 Python solutions from the aforementioned Github page,  
filter out 347 solutions\footnote{They are not self-contained, i.e. they cannot run as standalone programs.}, and end up with 1067 Python programs, each  with a few available unit tests. 
For comparison reasons, we translate these programs with the other transpilers too and report the results in Table~\ref{tab:caseleet}. 
JavaScripthon and py2js cannot translate any of the programs because they rely on the latest syntax and APIs (Python 3.9) which are not supported by these two transpilers. 
Transcrypt and Codex translate around half of the programs. 
\tool initialized by the 142 rules for GFG benchmarks cannot translate any program, however, once we add 16 rules its accuracy quickly surpasses Transcrypt and Codex. By adding more rules, it achieves even higher accuracy: 61.4\% with additional 41 rules, and 75.4\% with 106 rules. 
Its  average translation time is 1.75 seconds per program.

The above JavaScript translations provided by \tool are based only on a few available offline unit tests. To make sure the translations pass an actual LeetCode online assessment,  
we randomly sample 100 JavaScript programs from the 805 offline-correct translations and manually submit them to LeetCode. 
The site reported that 95 out of 100 passed all of the internal unit tests. 
This implies that with confidence of 98\%, the online accuracy on all programs is ($71.6 \pm 4.7$) \%, thus much higher than even the offline accuracy of any other compared transpiler. 
We can speculate that the close match of online to offline accuracy occurs because the incremental paradigm results in a search space that is small enough and highly constrained by the source, hence several examples might already be sufficient to filter out all the wrong translations in this space.

\subsection{Case Study: Translating longer code
}
\label{sec:casectci}

We test the scalability of \tool by translating longer programs. For this purpose, we use   Python programs\footnote{Programs were collected from the book's \href{https://github.com/careercup/CtCI-6th-Edition-Python}{official GitHub repository}.} from the book "Cracking the Coding Interview"~\cite{mcdowell2015cracking}. 
We collect all programs with more than 50 lines of code, in total 25 programs, of which
19 have 54 -- 100 lines and 6 have 100 -- 160 lines. 
On average the 25 programs have 88 lines (cf. to 15 lines of LeetCode programs), and some of them have multiple functions and class definitions. The overall line coverage of the unit tests is 88\%. 

As before, we start \tool with 142 rules obtained from the GFG benchmarks, and gradually add new rules. The final results and comparison to other transpilers are shown in Table~\ref{tab:caseleet}. 
With 197 (142 + 55) rules, \tool translates 14 out of the 25 programs. 
On the other hand, Codex is only able to translate three programs.
\tool needs 6.78s on average to search for a translation that passes tests, which is expected, since the longer the code, the more options \tool needs to check to find the correct translation. The longest time is 25s for \tool to translate a 114-line Python program. At the same time, Codex needs 37.6s on average because of the large number of tokens to decode. Besides \tool and Codex, Transcrypt is able to translate 11 of the 25 programs, but the translation is only emulating Python's classes into bloated JavaScript wrapper functions which are not native JavaScript classes thus sacrificing readability and maintainability. 

Since the test coverage is 88\%, we also manually check the translation of the uncovered parts of the code. It turns out that the remaining 12\% are mostly functions for debug printing, overridden function stubs and occasionally unused dead code. While not affecting the functionality of the translation, \tool cannot automatically fix errors  in the 12\% uncovered parts.

\end{todoenv}

\begin{todoenv}
\section{Discussions}
\label{sec:discussions}

Let us take a look at 
application domain, comparison with compilers, and extensibility of \tool.

\subsection{\tool Application Domain and Usage Model}
\label{sec:usagemodel}

Code translation in different application domains can vary drastically in terms of challenges and techniques. For use cases like porting code from Python to JavaScript, the translation requires moving across execution environments (e.g., from Python VM to node.js runtime) and so the sets of available APIs and libraries are not the same. This can be more challenging than translating within the same execution environment, such as CoffeeScript to JavaScript (within node.js), Visual Basic to C\# (within .NET), and Kotlin to Java (within JVM). In the latter cases, when available APIs and runtime behaviors are not changing, the translation is mostly focused on language features rather than data types and APIs, thus compilers with a small set of rules might work well for such tasks. On the other hand, for code migration across execution environments and dependencies, existing tools are mostly ad-hoc or less accurate. Examples of transpilers from both application domains are shown in Table~\ref{tab:appdomain}).

\begin{table}[ht]
\caption{Example code translation tools for different application domains} 
\label{tab:appdomain}
\centering
\small
\begin{tabular}{|l|l|}
\hline

\underline{across env.}

& \begin{tabular}{@{}ll@{}} 
    e.g., \underline{\tool}, TSS\footnotemark, Transcrypt, Codex, TransCoder\\ 
  \end{tabular}

\\

\hline

same env.

& \begin{tabular}{@{}ll@{}}  
    e.g., TypeScript / CoffeeScript (node.js), .NET Reflector (.NET), Fernflower (JVM)\\
  \end{tabular} 

\\
\hline
\end{tabular}
\end{table}
\footnotetext{Tangible Software Solutions: Source Code Converters, \url{https://www.tangiblesoftwaresolutions.com/converters.html}}

The usage model of \tool includes applications in line with Section~\ref{sec:caseleet}, i.e. to achieve availability of JavaScript code from Python code. 
We emphasize that the translation rules of \tool need to be created once if no customization is required, and can be used in all further translations. Thus the difficulty barrier of using the tool is low. 
For instance, \tool can be used to translate programming challenges for a tutorial website. The translation rules can be created once by the developer of the website. The end-users familiar only with some languages can use \tool to automatically translate solutions from other languages, unfamiliar to them. So, from this example, we can see that the rule creator and the end user do not have to be the same entity. 

When the unit tests are insufficient, additional correctness oracles might be needed. 
In the aforementioned example, if the users are not confident about the correctness of the translation, they can submit it to the tutorial website online judge system to get feedback, and in case of reported failure, they can run \tool with additional unit tests.

\subsection{Comparing \tool's Translation Rules With Rule-based Transpilers}
\label{sec:comparecompiler}
Traditional transpilers may differ in the used data structures and abstraction of rules (see detailed discussions in supplementary materials). However, 
it is common that their translation processes are driven by manually-specified state machines
that analyze the code and check the applicable conditions of transformation rules. 
Such state machines are typically implemented as visitors\footnote{\url{https://en.wikipedia.org/wiki/Visitor\_pattern}} of AST trees. 
In contrast, \tool's translation process is driven by grammar and unit test feedback rather than hand-crafted code analysis. Unlike the other transpilers whose rule compositions depend on opaque internal states of AST visitors, \tool has transparent behaviors because rule compositions are not associated with any manually-specified state change. 
To adapt to the context and to reduce errors, \tool adaptively reverts incorrect rule compositions based on the feedback from the test-driven loop, rather than using a manually-written deterministic decision procedure like compilers.

\subsection{Extending to Other Languages}
The current design of \tool can be adapted 
to language pairs with common memory management, data types, and programming paradigms. For example, Python, JavaScript, Lua, and ActionScript have similarities in those three aspects, despite 
targeting different execution environments with different APIs. 
To extend \tool to other pairs of languages, one needs to provide a tree-sitter grammar\footnote{The tree-sitter grammar might need some adjustments if external scanners are used.} (already available for many popular languages) and a pretty printer for  language's tree-sitter AST. 

To extend to less similar language pairs, we speculate that the current DSL and purely syntax-based solutions are not sufficient. For example, for a pair of a dynamically-typed and a statically-typed language, \tool may struggle to produce a type-consistent translation, or correct translations may be out of the search space if the code is JIT-unfriendly (one variable is assigned to many different types at runtime \cite{gong2015jitprof}). Another challenging case is translating languages with garbage collection (such as Go) to languages with manual memory management (such as C/C++). Without pointer analysis, \tool may introduce memory errors. 

\end{todoenv}
\begin{todoenv}
\subsection{Limitations and Future Work} 
\label{sec:limitations}

Let us summarize \tool's limitations and take a glance at potential future work.

\noindent \textbf{Insufficient unit tests.}
When the unit tests provide only partial code coverage, \tool with the test-based correctness oracle
is not able to get test feedback on the uncovered code lines to repair errors. 
Companies such as Google provide sufficient unit tests~\cite{ivankovic2019code} (80\% \textasciitilde 90\% on average), however, this is not always the case for open-source projects. While some popular Python projects have 90\% coverage \cite{zhai2019test}, for other codebases it is more difficult to write unit tests (e.g., for user interface code) \cite{fard2017javascript}. 
To tackle the problem of insufficient unit tests, one may use automated test generators to increase the coverage. An alternative is to use static analyzers and formal verification tools as correctness oracles to provide additional feedback to \tool, however, for dynamically-typed languages, such techniques might be limited, especially when the input specifications are not available. In the worst case, the user might need to review the uncovered part of the translation and manually choose the correct rules. 

\noindent \textbf{Differences in execution environments.} 
Execution environments may exercise different behavior on similar types. E.g. 
when iterating over a dictionary, Python virtual machine may use insertion order, but the order of JavaScript's node.js runtime may not be predictable~\footnote{\url{https://stackoverflow.com/questions/5525795/does-javascript-guarantee-object-property-order}}. So, if Python code depends on the iteration order, it cannot be translated straightforward.
The solution adopted by Transcrypt is to emulate Python's data types in JavaScript, but it may not be suitable for code migration due to non-idiomatic code and performance penalties. 

\noindent \textbf{Expressiveness of DSL and human effort.} The rule DSL of \tool, which directly maps to tree transducer, cannot match on sub-tokens inside leaf AST nodes. For example, it  cannot translate \code{"score: \%d"} into \code{"score: \{\}"} in a general way because \code{"\%d"} is inside the leaf node. It also cannot perform complex computations such as reordering keyword arguments in a function call based on the function's definition. A temporary solution without modifying the system is to use one-shot rules for those corner cases, but such a trick will not reduce human effort and will result in many non-reusable rules. A more expressive rule DSL might support more complex transformations but rule inference might also become more challenging. We leave as an open problem the design of better translation DSL with rule inference that may further minimize overall human effort.

\noindent \textbf{Scalability.} When the code is longer, as shown in the CtCI case study from Section~\ref{sec:casectci}, all  Python to JavaScript transpilers have lower accuracy due to accumulated translation errors and corner cases. 
Specifically, for \tool 
half of the failed translations from CtCI contain corner cases that cannot be automatically fixed, while for the correct translations the number of retries increases (e.g, up to 25 seconds and 52 retries for 114 lines of code). 
So, longer code might need to be split into sub-tasks for \tool to handle it efficiently. 
In some extreme cases, when the running time of the program is high (e.g. so-called \code{pytest} unit testing framework), despite the small number of retries, \tool would require a larger amount of time to run.
Better divide-and-conquer strategy, smarter rule selection, or faster correctness checks may increase \tool's scalability and all of these constitute compelling future work.

\end{todoenv}
\section{Related Work}
\label{sec:related}

There are a few broad approaches used for code translations that we summarize below.

\paragraph{Statistical and neural machine learning based translation} Various statistical machine translation techniques have been proposed. \cite{karaivanov2014phrase}, \cite{nguyen2013lexical} and \cite{nguyen2015divide} use phrased-based statistical translation. \cite{graehl2008training} proposed a training algorithm for tree transducers.  \cite{nguyen2016mapping} use statistical methods to find mappings between APIs. Recently, neural-based translators have been gaining in popularity. \cite{xinyun2018tree} use tree-to-tree neural network to translate JavaScript to CoffeeScript. \cite{roziere2020unsupervised} proposed TransCoder that learns to translate code using unsupervised learning. Besides the specialized translators, large language models (such as Codex~\cite{chen2021evaluating}) are trained for many code-related tasks, including translation. However, all of the statistical methods require a large amount of data to train. In our approach, we decouple the syntactic and semantic aspects of the translation. This allows us to learn customized syntactic patterns from few short code snippets and search for semantically correct translation on the fly.
\paragraph{Term-rewriting systems} Conventionally, source code translation can be accomplished with manually written rule-based term rewriting systems and pattern matching. StringTemplate\footnote{\url{https://www.stringtemplate.org/index.html}} is a popular framework for rule-based code generation and translation.  TXL~\cite{cordy2006txl} is a general-purpose source transformation language based on union grammar techniques and manually-written transformation rules. SrcML~\cite{collard2013srcml} is another system for lightweight source transformation within the same language. Some computer algebra systems (such as Mathematica~\cite{wellin2005introduction}) have expressive pattern language to transform mathematical expressions between alphabets or to export to \LaTeX. All these transformation tools require experts to manually write the rules, and syntactic or semantic correctness are typically challenging when translating between languages with different grammar and semantics.

\paragraph{Program Synthesis} An alternative approach is to use program synthesis to solve code translation and transformation problems.  \cite{kamil2016verified} proposed verified lifting to synthesize a semantically equivalent parallel code from a piece of low-level sequential code based on SAT/SMT solving.  \cite{mariano2022automated} propose a neural-guided program synthesizer with similar functionalities using cognate grammar network. However, only short programs can be transformed this way due to scalability issues in typical program synthesizers.   \cite{rolim2017learning},\cite{miltner2019fly} synthesize short code-transformation programs doing automated code refactoring for several statements within the same language, which is orthogonal to our work as we translate complete programs across different languages.

\section{Conclusion}
\label{sec:conclusion}

We propose a new design for transpilation with a focus on democratizing the process of building and using such transpilers. With our design, the end user would be able to build custom transpilers that are tuned to their code style, different APIs and preferred language features with little manual effort. The key design principle is to incrementally learn the rules that are only required to translate the program at hand from user-provided code snippets. Through the snippets, the users may fully customize the translation rules and tailor such rules to their needs. The manual effort is also limited to specifying a few short code snippets only once per rule that can then be reused in the future. Internally, the design formalizes the translation problem as a search in a program space structured by non-deterministic tree transducers, customized by transduction rules, and constrained by grammar and unit tests. 

We have also provided a practical instantiation of our design called \tool that translates from \pylang to \jslang. \tool learns \totalrulecount rules from $360$ lines of snippets to reach $90\%$ accuracy on GeeksForGeeks benchmarks, which is of \benchlc lines in total ($14.7\times$ of the code snippets). As a comparison, none of the other transpilers achieves an accuracy higher than $76\%$. Similarly, none provides flexible customizability or non-expert upgradability of translation rules. \tool can be fairly easily adapted to translate between other untyped scripting languages.

\bibliographystyle{ACM-Reference-Format}
\bibliography{paper}

\end{document}

\title{Supplementary Material: User-customizable Transpilation of Scripting Languages}
\maketitle

\appendix

\section{Optimizing Translation Search}
\label{sec:ds}

\begin{figure}[ht]
    \centering
    \includegraphics[width=0.7\linewidth]{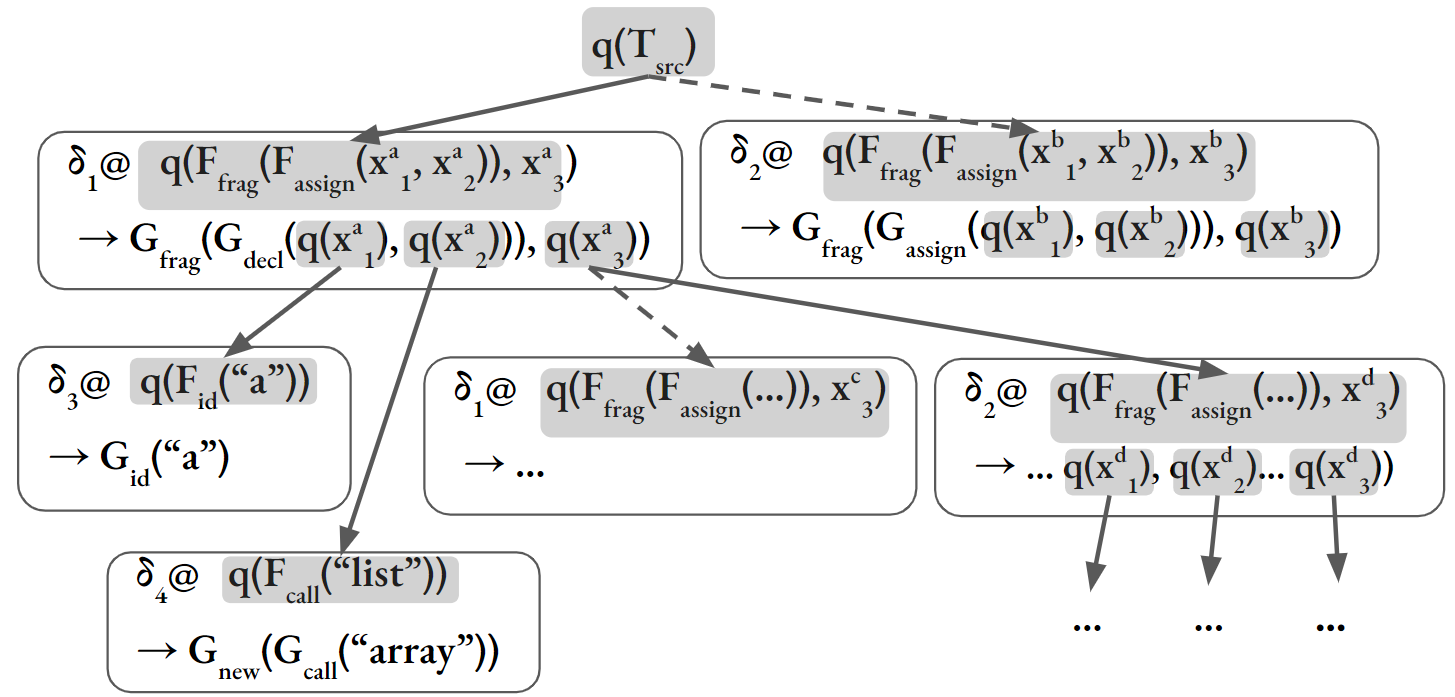}
    \caption{Non-deterministic derivation tree is lazily constructed as rules are selected. The solid arrows point to selected transductions and the dashed arrows are not selected ones of which further transductions are delayed. $\delta_1 @ q(\cdots)$ means the transduction step of applying the rule $\delta_1$ on the subtree in $q(\cdots)$'s form. }
    \label{fig:lazytrans}
\end{figure}
\begin{figure}[ht]
    \centering
    \includegraphics[width=0.6\linewidth]{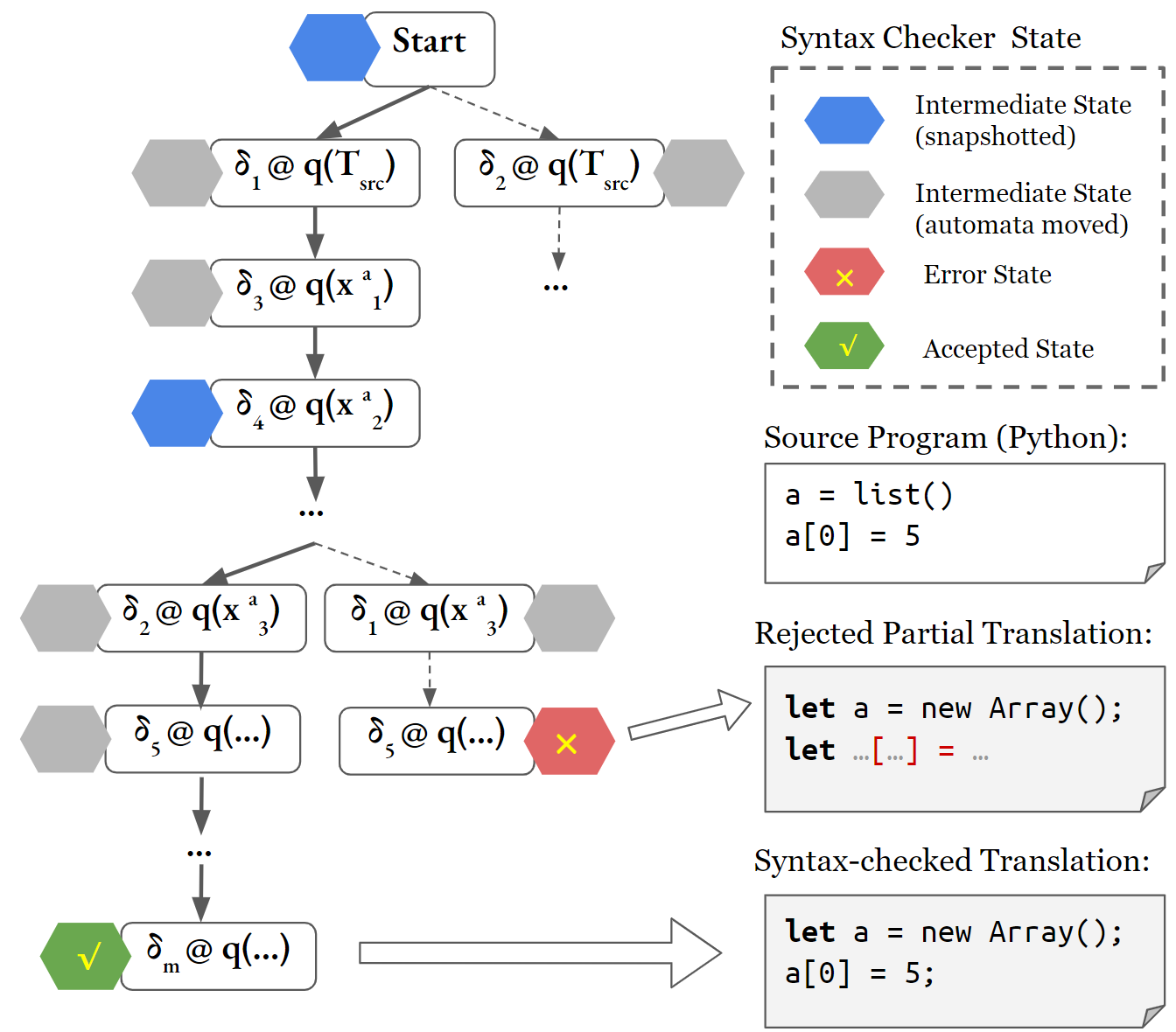}
    \caption{The leftmost derivation history tree. Each path in the history tree is a sequence of transduction rules used to do top-down depth-first transduction. The solid and dashed arrows in figure \ref{fig:lazytrans} correspond to solid and dashed arrows in this figure. Each path in the deriviation history is validated by a syntax checker (an automaton). The blue shape in the figure is a snapshotted automaton that and can be cloned later to reduce the computation cost of exploring different paths from the snapshotted node.}
    \label{fig:transhistory}
\end{figure}

To aid the search for a correct translation, \tool uses two lazily-evaluated data structures: a non-deterministic derivation tree, and a leftmost derivation history tree. These data structures are instrumental to the syntax and semantic checkers.

The non-deterministic derivation tree (Figure~\ref{fig:lazytrans}) is a tree representing all possible transductions on a specific source AST, i.e., when all possible transductions for every node in the tree are fully expanded. This tree is used to search for a possible rule to apply when a source AST is being translated. There are two types of nodes in this derivation tree: transduction nodes and slot nodes. The root node is a slot node representing the task of transducing the whole source AST, i.e., $q(T_{\text{src}})$. Each node in the transduction tree has multiple possible transduction nodes as its children where each one represents a rule (see round boxes) with further slot nodes. The function-like symbols using letters $F_{\text{xxx}}$ and $G_{\text{xxx}}$ in the round corner boxes are multi-level partial trees and their arguments are the subtrees of those partial trees. For example, $F_{\text{assign}}(x^a_1, x^a_2)$ is the partial AST representing an assignment and $x^a_1$ and $x^a_2$ are the two sides of the assignment. Applying a transduction rule results in a target partial AST with multiple slots $q(x_1), q(x_2), \ldots$ representing subtasks created by the transduction rule. The tree is constructed lazily to make the search operations on the tree efficient. 

The leftmost derivation history tree (Figure~\ref{fig:transhistory}) is a tree representing all possible leftmost derivations of the transducer (when fully expanded). Each path represents the transduction history of one leftmost derivation.
We use a non-deterministic pushdown automaton with instruction logs for validation and parse tree reconstruction. In the leftmost derivation tree, each path is validated by our automaton. In Figure~\ref{fig:transhistory}, each arrow is one step of the leftmost derivation validated by the automaton from the previous step. Normally, the automaton has state changes after each validation step and moves along the arrows. To reduce the computational cost of exploring different transductions (corresponding to different derivation histories in this tree), the automaton's stack and its instruction logs are using immutable data structures for fast snapshotting. The blue shape in Figure~\ref{fig:transhistory} is a snapshotted automaton so that all derivations from that point can clone from the automaton snapshot to validate different derivations. As an example in the figure, two different derivations are explored from the snapshot. The right path is applying several transduction rules to translate the second line of \pylang source program (line \code{a[0] = 5}). The first rule $\delta_1$ transduces \pylang assignment \verb|. = .| to \jslang declaration \verb| let . = .;|. The second rule translates \pylang subscript operator \verb|.[.]| to \jslang subscript \verb|.[.]|. However, at this point, the derivation is rejected because the automaton enters the error state after accepting the second transduction. The reason is that in \jslang grammar, the LHS of the declaration must be an identifier. The other derivation on the left passes the automaton's validation and a valid parse tree is reconstructed from the automaton's instruction log (bottom right in Figure~\ref{fig:transhistory}).

The leftmost derivation history tree is also lazily constructed and validated on the fly as rules are selected. Different paths in the derivation history tree might share many transduction steps in the non-deterministic derivation tree. To search for syntactically valid translations, \tool will search for paths in the leftmost derivation history tree that is accepted by the validating automaton. \tool has default preferences over rules. The default behavior is that rules with more specific LHS (closer to terminals) have higher priority than rules with more general LHS. For rules with the same LHS, the rule that comes first in the file for the ruleset is preferred.

\section{Details of Rule Inference Algorithm}

The SimultaneousTraversal algorithm used by the rule inference algorithm is shown in Algorithm~\ref{alg:simudfs}.

\begin{algorithm}[t]
    \caption{Simultaneous Traversal Algorithm}
    \label{alg:simudfs}
    \begin{algorithmic}[1]
      \Function{NodesVisitor}{nodes}
        \If{AllEqual(nodes.map(node $\rightarrow$ node.name))}
            \If{AllEqual(nodes.map(node $\rightarrow$ getChildCount(node)))}
              \If{AllTrue(nodes.map(node $\rightarrow$ isTerminal(node)))}
                \If{AllEqual(nodes.map(node $\rightarrow$ node.value))}
                  \State \Return nodes[0], True
                \Else
                  \State terminalType = isFixedStr(node) ? "\_str\_" : "\_val\_"
                  \State diffPH = \textbf{createTplPH}(terminalType, nodes.map(node $\rightarrow$ node.value))
                  \State \Return createNode(nodes[0].name, children=[diffPH]), True
                \EndIf
             \Else
              \State \Return createNode(nodes[0].name), False
             \EndIf
            \Else 
              \State diffPH = \textbf{createPH}("*", nodes.map(node $\rightarrow$ node.children))
              \State \Return createNode(nodes[0].name, children=[diffPH]), True
            \EndIf
        \Else
            \State \Return \textbf{createPH}(".", nodes), True
        \EndIf
      \EndFunction
      \Procedure{SimultaneousTraversal}{ASTFragments}
        \State commonNode, isFinished = NodeVisitor(ASTFragments)
        \If {isFinished}
          \State \Return commonNode
        \Else
           \State nodes\_groups = zip(ASTFragments.map(node $\rightarrow$ node.children)) 
           \State commonNode.children = nodes\_groups.map(nodes $\rightarrow$ SimultaneousTraversal(nodes))
           \State \Return commonNode
        \EndIf
      \EndProcedure
    \end{algorithmic}
\end{algorithm}

\section{Details of \tool's User Interface}

\begin{figure}[ht]
    \centering
    \includegraphics[width=0.8\linewidth]{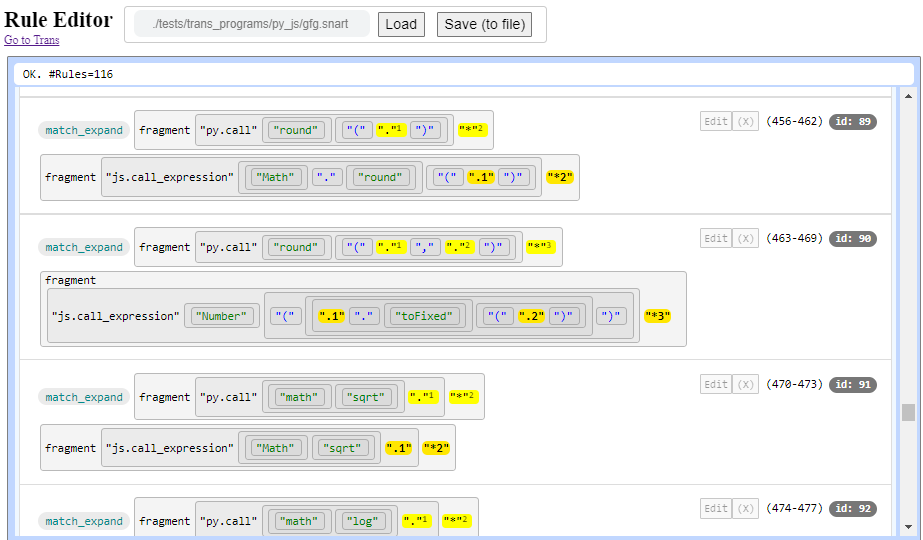}
    \includegraphics[width=0.8\linewidth]{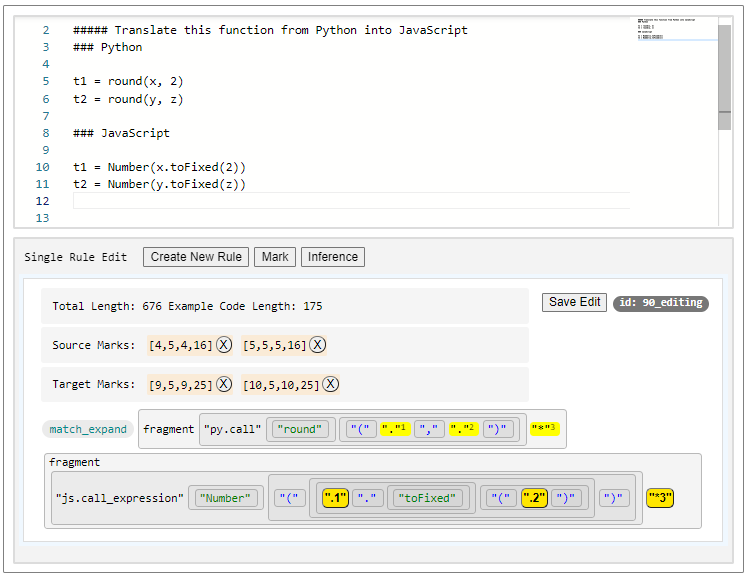}
    \caption{User Interface for Rule Editing and Inference}
    \label{fig:demoruleeditor}
\end{figure}

\begin{figure}[ht]
    \centering
    \includegraphics[width=\linewidth]{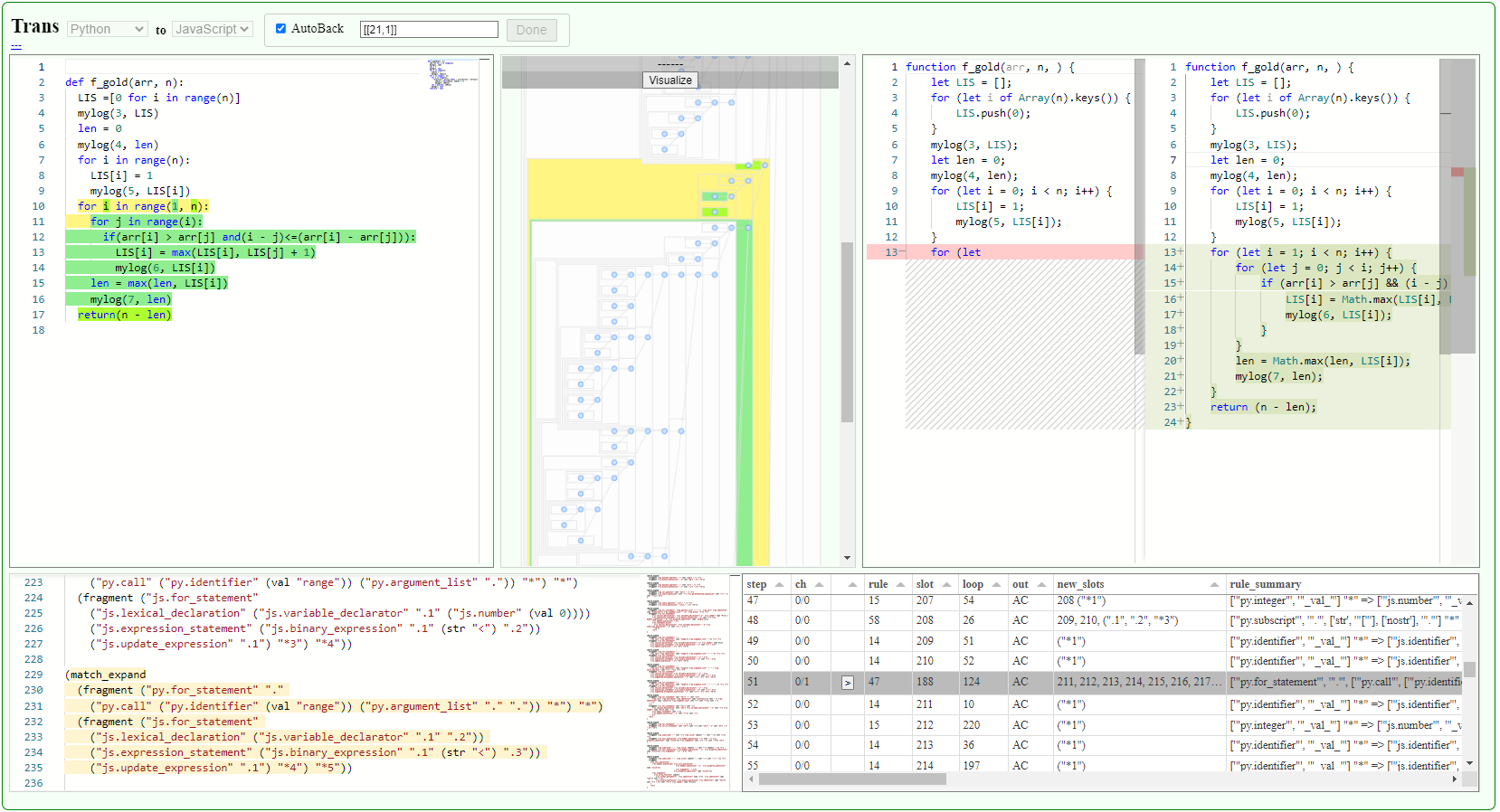}
    \caption{User Interface for User to Inspect and Altering the Translation. In the figure, the interface is showing rule mapping of the translation of \texttt{for i in range(1. n) ...} to \texttt{for(let i = 0; i < n; i++) ...}}
    \label{fig:demotransui}
\end{figure}

A screenshot of the rule editor of \tool is shown in Figure~\ref{fig:demoruleeditor}. 
A screenshot of the Translation UI for user to inspect or alter the translation is shown in Figure~\ref{fig:demotransui}. 

\input{figtabs/tab-percentdiff}

\section{Translation-uniqueness Comparison}
We also evaluate the additional translation uniqueness that \tool brings to the world of transpilers. More precisely, we
estimate the percentage of benchmarks that can be translated only by \tool, but not by other transpilers, and vice versa. We present two settings: one-vs-one and one-vs-all, refer to Table~\ref{tab:diffsolve}. 
In the one-vs-one setting, \tool can translate from $18\%$ to $69\%$ more benchmarks compared to the other transpilers, while it 
fails at most at $2.4\%$ of the benchmarks. 
Moreover, even in the one-vs-all setting, \tool still performs better than all four  transpilers together: it can translate $5.4\%$ additional benchmarks, while lacks in $3.6\%$.

\section{Detailed discussion of transformation rules in traditional compilers}
\noindent \textbf{Translate within the language.} Compiler tools in this categories usually perform a sequence of scoped edits to transform the source code into target code, such are TypeScript compiler and Clang AST Transformer. TypeScript is designed to be a superset of the newest version of JavaScript so that the compiler uses a chain of passes to transform the code to an older version (a subset) of the source language (TS -> ESNext -> ES2022 -> ES2020  ...). Each of those passes has a stateful visitor maintaining context information across different nodes. Such a compiler is not designed for customization, because the developer need to understand the interactions between node visiting functions and the visitor's context and also low-level details about how to match and create AST nodes. To address those issues, Clang AST Transformer has a high-level description language to write independent transformation rules that do not need to interact with a stateful visitor. Each of the rules has a matcher made of nested nouns (names of AST Nodes) and predicates (properties of nodes computed by Clang) to match on some AST nodes, and a replacement pattern that evaluate to the replacing code string. However, as mentioned in their online tutorial, to be able to accurately match the intended code segment, the developer not only needs to be familiar with Clang's AST structure, but also need to manually write appropriate predicates while their are hundreds of such predicates to choose from. When the description language is not precise enough for a given task, developer need to manually write filtering conditions to further limit the scope of rules. While \tool is not designed for in-language transformation, it is worth noting that the developer does not need to write those predicates and filtering conditions. It is also unnecessary for the developer to know the compilers' AST structure to create rules.

\noindent \textbf{Translate within the execution environment.}  Compiler tools in this category usually have an intermediate representation (IR) that can be transformed from the source language and pretty-printed to the target language. The translation can usually guarantee the same semantics from the execution environment's perspective (language runtime or OS). Some examples are CoffeeScript (compiles to JS, both for browser), CodeConverter (VB to C\#, both on .NET), and Cython (Python module to Python-loadable C module). The CoffeeScript compiler will desugar and normalize the AST so that it directly map to semantically-equivalent JavaScript for pretty printing. The CodeConverter uses Roslyn compilers IR to bridge C\# and VB, because Roslyn compiler is using a shared IR for those two languages. The Cython compiler translate Python code to equivalent C code that calls to Python interpreter to create, load, store and operate Python objects. The translated C module, when loaded by Python interpreter, has the same semantics as interpreting the original Python code, and it cannot run without the Python interpreter. Transformation rules in those translators are similar to in-language transformaton tools, which interact with context to compute application conditions, and every performed transformation step should preserve semantics.  Extra care is also needed to make sure that pretty-printing in the final step results in syntactically valid code in the target language.

\noindent \textbf{Translate across the execution environment.} Compiler tools in this category are usually far from satisfying. Due to the difficulty of defining shared semantics across different execution environments, designs for the previous 2 categories is less successful.  Some translators (such as Transcrypt) mix emulation and translation. They attach a library to the translated code that emulates builtin APIs and operators of the source execution environment. This approach is limited because emulating the source execution environment will either have huge performance overhead or require significant engineering effort. Some customizable translators such as TXL use union grammar to model the translation as a sequence of piece-by-piece transformations of the source, similar to Clang AST Transformer. However, the union grammar of two languages is difficult to write and its tricky to define semantic-preserving transformations because a simple syntactic mixture of code written for two different execution environments is not interpretable. While syntax-level rewriting patterns are easy to write in TXL, the conditions for applying those rules and compositions of them are subtle and error prune.